\documentclass{article} 
\usepackage{iclr2024_conference,times}

\usepackage{amsmath,amsfonts,bm}

\def\eqref#1{equation~\ref{#1}}

\def\1{\bm{1}}

\DeclareMathAlphabet{\mathsfit}{\encodingdefault}{\sfdefault}{m}{sl}
\SetMathAlphabet{\mathsfit}{bold}{\encodingdefault}{\sfdefault}{bx}{n}

\DeclareMathOperator*{\argmin}{arg\,min}

\usepackage{hyperref}
\usepackage{url}
\usepackage{multirow}
\usepackage{subcaption}
\usepackage{wrapfig}
\usepackage{graphicx}
\usepackage{xcolor}
\usepackage{todonotes}
\usepackage[most]{tcolorbox}
\usepackage{amsmath, amssymb}
\usepackage{pgfplots}
\usepackage{tikz-3dplot}

\usepackage[inline]{enumitem}

\newtcolorbox{definition}[1][]{
    colback=blue!10,  
    colframe=blue!10!black,    
    fonttitle=\bfseries,
    sharp corners,            
    breakable=false, 
    title=#1
}
\newtcolorbox{example}[1][]{
    colback=gray!10,   
    colframe=gray!10!black, 
    fonttitle=\bfseries,
    breakable=false,          
    sharp corners,            
    title=#1
}

\title{Equivariant Amortized Inference of Poses for Cryo-EM}

\author{Larissa de Ruijter \thanks{Work done during internship at QUVA Lab.} \\
QUVA Lab, University of Amsterdam \\
\texttt{larissaderuijter@kpnmail.nl} \\
\And
Gabriele Cesa \\
Qualcomm AI Research, Amsterdam\thanks{Qualcomm AI Research is an initiative of Qualcomm Technologies, Inc.} \\
QUVA Lab, AMLab, University of Amsterdam \\
\texttt{gcesa@qti.qualcomm.com} \\
}

\iclrfinalcopy 

\begin{document}

\maketitle

\begin{abstract}
Cryo-EM is a vital technique for determining 3D structure of biological molecules such as proteins and viruses. The cryo-EM reconstruction problem is challenging due to the high noise levels, the missing poses of particles, and the computational demands of processing large datasets. A promising solution to these challenges lies in the use of amortized inference methods, which have shown particular efficacy in pose estimation for large datasets. However, these methods also encounter convergence issues, often necessitating sophisticated initialization strategies or engineered solutions for effective convergence.
Building upon the existing cryoAI pipeline, which employs a symmetric loss function to address convergence problems, this work explores the emergence and persistence of these issues within the pipeline. Additionally, we explore the impact of equivariant amortized inference on enhancing convergence. Our investigations reveal that, when applied to simulated data, a pipeline incorporating an equivariant encoder not only converges faster and more frequently than the standard approach but also demonstrates superior performance in terms of pose estimation accuracy and the resolution of the reconstructed volume. 
Notably, $D_4$-equivariant encoders make the symmetric loss superfluous and, therefore, allow for a more efficient reconstruction pipeline. 

\end{abstract}

\section{Introduction}
Cryo-electron microscopy (cryo-EM) has emerged as an crucial technique in molecular biology and chemistry, enabling the determination of macro-molecular structures such as proteins. 
In cryo-EM, particle samples are frozen in a thin layer of vitreous ice and exposed to an electron beam. The interaction between electrons and the sample's electrostatic potential scatters electrons in patterns that reflect the molecular structure. This results in multiple noisy two-dimensional projections of the particles in random orientations, which are then used to reconstruct the molecular structure. 

The reconstruction process presents significant challenges, including the estimation of unknown poses and the high noise levels \cite{singer2011three,singer2020computational}. 
Traditional methods employ iterative approaches that, in each step, require extensive searches over the space of poses \emph{for each datapoint} and a refinement of the molecular structure estimate (e.g. maximum-likelihood expectation-maximization (ML-EM) implemented by SOTA softwares like RELION \cite{scheres2012relion} and cryoSPARC \cite{punjani2017cryosparc}); 
however, this per-image search 
doesn't scale well with the increasing datasets sizes due its computational demand \cite{levy2022cryoai}. 

Recently, deep learning approaches have shown promise in addressing these challenges, giving rise to a new category of reconstruction methods based on \textit{amortized inference} \cite{giri2023deep}. These methods learn a parametrized function, typically within a framework employing (variational) autoencoder architectures, to predict particle poses from images, shifting the computational burden to the learning phase. Although this reduces the need for per-image pose estimation, these solutions often face convergence issues, requiring careful engineering to achieve~satisfactory~performance.

\paragraph{Contributions}
We introduce \textit{equivariant} amortized inference of poses for cryo-EM. This method uses the property that images that differ by some in-plane rotation or mirroring have associated poses that undergo a similar transformation. 
In the amortized inference framework,
we can exploit this prior knowledge of the geometry of the problem by making the parameterized function \emph{equivariant} with respect to rotations and/or reflections.
Using an equivariant function has the following benefits:
\begin{itemize}
    \item \textit{faster generalization $=$ faster reconstruction}: once it learns to predict the pose of an image correctly, it immediately generalizes to rotations and reflection of that particular image. This typically also leads to faster convergence and, thus, to faster reconstruction here.
    \item \textit{geometrical consistency = reduced convergence issues}: it is forced from the start to represent subsets of datapoints according to the correct underlying geometric structure, potentially reducing or preventing the convergence issues discussed earlier. 
\end{itemize}

\section{Image formation model}
\label{mathcryo}
We can model the electrostatic potential of a molecule as a 3D density, \(V\colon\mathbb{R}^3\to\mathbb{R}\). Each molecule, frozen in ice, assumes a random orientations represented by a rotation \(R_i \in \text{SO}(3)\). The image is formed by electrons that are scattered by interacting with this potential, hitting a detector. This is modeled as a projection:
\begin{equation}
    P_{R_i}[V]\colon (x,y) \to \int_z V(R_i^{\top} (x,y,z)^{\top})\,dz.
    \label{eq:P_i}
\end{equation}
Before the electrons hit the detector, they interact with the lens system of the microscope, represented by a convolution with a Point Spread Function (PSF) kernel \(h_i\). Additionally, we account for small translations \(t_i\in \mathbb{R}^2\) by convolving with the translation kernel $T_{t_i}$ and add a Gaussian noise variable \(\epsilon_i\) to model various noise sources. The resulting image formation model is:
\begin{equation}
    I_i= h_i \ast T_{t_i} \ast P_{R_i}[V] + \epsilon_i,
\end{equation}
We assume square integrable functions, i.e. $P_{R_i}[V]$ and $I_i\in L^2(\mathbb{R}^2)$.
The image formation model is usually formulated and implemented in Fourier space, which we describe in Apx.~\ref{app:im_fourier}.

\section{Related work}
\label{sec:rel_work}

\paragraph{Amortized inference of poses for cryo-EM}
Recently, deep learning based methods have been applied to the cryo-EM reconstruction problem \citep{donnat2022deep, giri2023deep}. With this, a new approach has emerged: amortized inference over poses. Such methods  learn a parametrized function that estimates a pose given an image, as opposed to estimating the pose for each image individually. Typically, (variational) autoencoder architectures are employed in this setting. For example, \cite{rosenbaum2021inferring} use a variational autoencoder architecture in a heterogeneous setting in which the encoder predicts both poses and conformations. CryoVAEGAN \citep{miolane2020estimation} is a pipeline in which an encoder estimates rotations as well as parameters of the contrast transfer function (CTF). In the spatialVAE pipeline the encoder estimates translations and in-plane rotations \cite{bepler2019explicitly}. CryoPoseNet \citep{nashed2021cryoposenet}, cryoAI \citep{levy2022cryoai} and cryoFIRE \citep{levy2022amortized} demonstrate the efficacy of an autoencoder (AE) architecture in which the encoder learns to estimate poses. While promising, amortized inference methods may suffer from convergence issues. For example, in \citep{rosenbaum2021inferring}, a strong prior on the backbone of the molecule is needed for the pipeline to converge. 
CryoPoseNet tends to get stuck in local minima while cryoAI and cryoFIRE limit this issue by using an additional symmetric loss.

\paragraph{Homoemorphic Encoders and Equivariance} 
\cite{falorsi2018explorations} introduced \textit{homeomorphic encoders} to describe a setting where the input data has a geometric structure associated with a symmetry group and the goal for the encoder is to learn a homeomorphic geometric representation of the data in the latent space.
This is analogous the amortized inference of poses in cryo-EM, as each image is identified by an element of $SO(3)$.
\cite{esmaeili2023topological} show that these models are prone to optimization challenges due to topological defects at initialization that cannot be resolved continuously, partially explaining the convergence issues observed in previous works employing amortized inference of poses.
One can incorporate geometric inductive bias into neural networks via \emph{equivariance}. 
For example, convolutional neural networks (CNNs) are translation-equivariant.
This idea is generalized via group convolution (GCNNs) \citep{cohen2016group} and Steerable CNNs \citep{cohen2016steerable} to achieve equivariance with respect to more general symmetry groups.
Cryo-EM images possess a number of symmetries \citep{cesa2022symmetries} which can be exploited via an equivariant design. 
Previously, \cite{nasiri2022unsupervised, pmlr-v221-cesa23a, granberry2023so} leveraged some of these planar symmetries.
To the best of our knowledge, this is the first work to successfully leverage these symmetries in an end-to-end cryo-EM reconstruction pipeline.

\section{Method}

\subsection{Equivariant amortized inference of poses}

We can think of a 3D rotation as an element $R_i=(x_i,y_i,z_i) \in SO(3)$, where $x_i,y_i$ and $z_i \in \mathbb{R}^3$ are the columns of the matrix $R_i$ associated to the rotation. 
The vectors $x_i, y_i$ and $z_i$ form an orthonormal basis of $\mathbb{R}^3$. The operator $P_{R_i}$ defined in Eq.~\ref{eq:P_i} first rotates the volume $V$ and then performs a
\begin{wrapfigure}[25]{r}{0.23\textwidth}  
    \centering
       \vspace*{-1.4em}
       \hspace*{-1.8em}
      \includegraphics[scale=0.35]{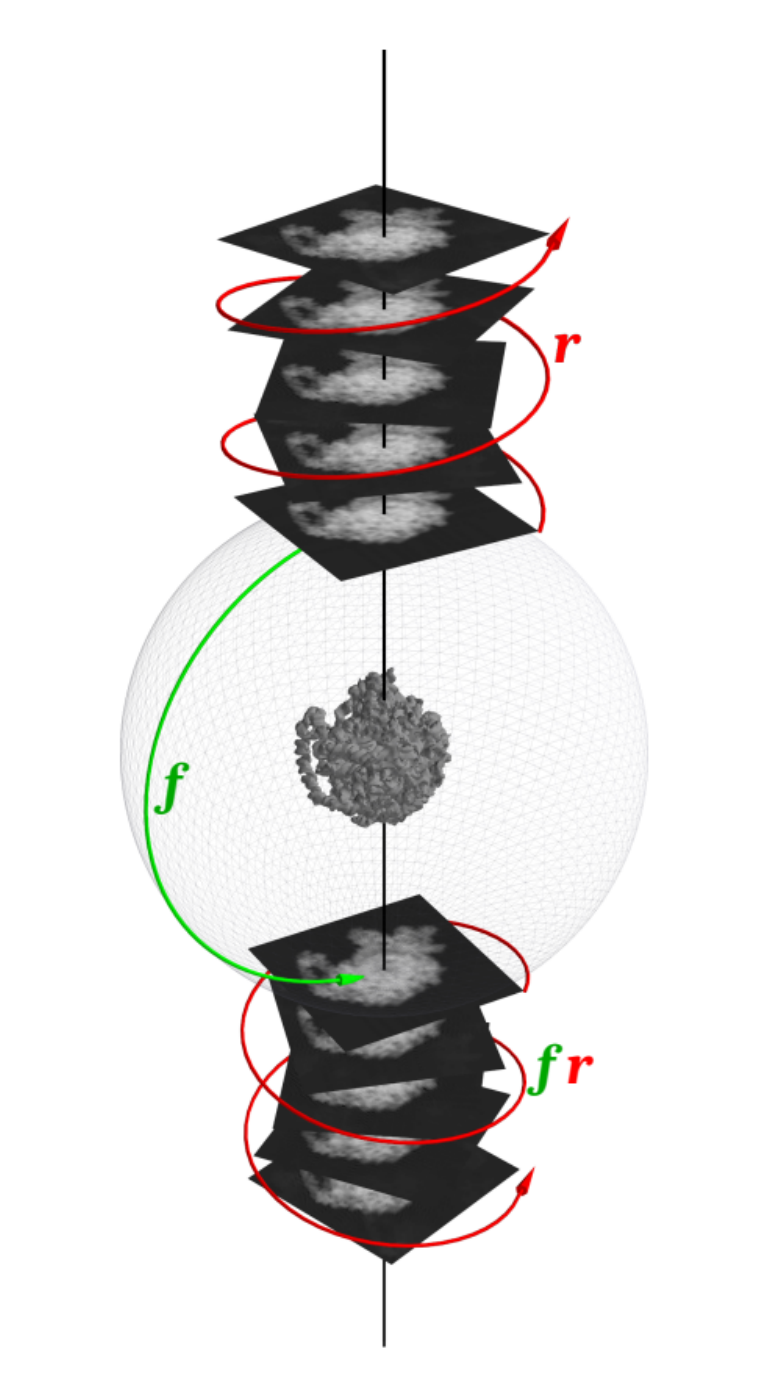}
       \vspace{-2.2em}
    \caption{\small Images in the same equivalence class differ only by an in-plane rotation and/or reflection. Source: \cite{cesa2022symmetries}.}
    \label{fig:equivariance}
\end{wrapfigure} 
projection along the $z$-axis. 
Rotating the volume $V$ by $R_i$ can be seen as transforming $V$ with the
transformation that would align $x_i,y_i$ and $z_i$ with the standard basis vectors of $\mathbb{R}^3$. Therefore, alternatively, we can see the axis $z_i$ as the `direction' along which the projection is performed. We shall call the axis $z_i$ the \emph{viewing direction}. 
Note that two images $I_i,I_j\in L^2(\mathbb{R}^2)$ that are generated by projecting along the same viewing direction $z$ can only differ by an in-plane rotation $r_{\theta}\in \text{SO}(2)$, whereas if they differ by a rotation and a reflection $f\cdot r_{\theta}\in\text{O}(2)$, they are obtained by projecting along opposite viewing directions $z$ and $-z$, see Fig.~\ref{fig:equivariance}. 
In fact, given an element 
\begin{equation}g= f^c\cdot r_{\theta}=\begin{bmatrix}1 & 0 \\0 & -1\end{bmatrix}^c\begin{bmatrix} \cos(\theta) & -\sin(\theta) \\ \sin(\theta) & \cos(\theta) \end{bmatrix}\in\text{O}(2)\label{eq:element_g}\end{equation}
with $c\in\{0,1\}\text{ and } \theta\in [0,2\pi),$ if two images $I_i$ and $I_j$ are related such that\footnote{\label{fn2}
    The \emph{left action} of $g\in \text{O}(2)$ on an image $f\in L^2(\mathbb{R}^2)$ yields a new image given by $[L_g[f]](x)=f(g^{-1}x)$.
}  $I_j=L_g[I_i]$, then we know that their associated poses $R_i,R_j\in \text{SO}(3)$ are related by the following identity:
\begin{equation}
    R_j=  \varphi(g)R_i=  \begin{bmatrix}
    -1 & 0 & 0 \\
    0 & 1 & 0 \\
    0 & 0 & -1
    \end{bmatrix}^c \begin{bmatrix}
    \cos(\theta) & -\sin(\theta) & 0 \\
    \sin(\theta) & \cos(\theta) & 0 \\
    0 & 0 & 1
    \end{bmatrix}R_i.
    \label{eq:3d_rep_o2}
\end{equation}
where $\varphi:\text{O}(2)\to \mathbb{R}^{3 \times 3}$ is a three-dimensional representation of O$(2)$; see Apx.~\ref{app:group_theory}.
Poses typically include also planar translations in $\mathbb{R}^2$, which follow an analogous equivariance rule; see Apx.~\ref{app:encoder_details} for more details.
In Sec.~\ref{subsec:equivariant_encoder}, we will exploit this knowledge 
in the amortized inference framework by restricting the parameterized function mapping images to their associated poses to be \textit{equivariant} to a subgroup $H\subseteq \text{O}(2)$.

\subsection{Pipeline overview}
\label{pipeline}
We build on the work by \cite{levy2022cryoai}, adapting their autoencoder pipeline by incorporating an equivariant encoder. 
Given an input image $Y$, the encoder $E\colon L^2(\mathbb{R}^ 2)\to \mathbb{R}^ 2\times \text{SO}(3)$ predicts a pose existing of a translation $t\in\mathbb{R}^2$ and a rotation $R\in \text{SO}(3)$.
The decoder maintains an estimate of the reconstructed volume via an implicit neural representation $\hat{V}\colon \mathbb{R}^3\to \mathbb{C}$, which models the Fourier transform of $V$.
The decoder simulates the image formation model in Fourier space by leveraging the \emph{Fourier Slice Theorem} as in Apx.~\ref{app:im_fourier}.
: it rotates a slice of three-dimensional coordinates $[k_x,k_y,0]^{\top}$ by the rotation $R$ and, then, feeds them into $\hat{V}$. 
The result is translated by $t$, convoluted with the contrast transfer function; see Apx.~\ref{app:decoder} for more details about the decoder.
Finally, the output $X$ is compared to the input $Y$ via a mean-squared error loss ($L2$ loss).

\subsection{Equivariant encoder}
\label{subsec:equivariant_encoder}
We replicate the architecture of the original encoder of cryoAI (see Apx.~B of \cite{levy2022cryoai}) using the \texttt{escnn} Python library \cite{cesa2022a} to incorporate equivariance to a subgroup $H \subset O(2)$; see Apx.~\ref{app:eq_enc_arch} for more details.
We can interpret the encoder $E$ as two separate functions: $E_r: L^2(\mathbb{R}^2) \rightarrow \text{SO}(3)$ that outputs a rotation, and $E_t: L^2(\mathbb{R}^2) \rightarrow \mathbb{R}^2$ that outputs a translation. 
Equivariance ensures that, for any element $g \in H$ and image $I$:
\begin{equation}
    E_r(L_g[I]) = \varphi(g)E_r(I),\,\, \text{ and }\,\, E_t(L_g[I]) = g\cdot E_t(I).
    \label{eq:rot_eq}
\end{equation}
In particular, we experimented with equivariance to the $H=C_4$ and $H=D_4$ subgroups of $O(2)$.
$C_4$ contains the rotations by $0,\frac{\pi}{2}, \pi, \frac{3\pi}{2}$, while $D_4$ contains these rotations, and their combinations with a reflection.
Note that these groups model the exact rotational symmetries of a pixel grid.

\subsection{Symmetric loss function}
\label{sec:loss}
The original pipeline is optimized using a \emph{symmetric loss function} to reduce convergence issues \cite{levy2022cryoai}. 
This loss function augments a batch with a duplicate of each image rotated by $\pi$.
For each image and its rotated duplicate, the model is only supervised on the one it reconstructs best.
Formally, the symmetric loss $\mathcal{L}_{\text{sym}}$ is defined as follows:
\begin{equation}
    \mathcal{L}_{\text{sym}}=\sum_{i\in B}\text{min} \{||\hat{Y}_i-\Gamma (Y_i)||^2,||L_{\pi}[\hat{Y}_i]-\Gamma (L_{\pi}[Y_i])||^2\},
    \label{eq:symloss}
\end{equation}
where $\Gamma$ is the whole cryoAI pipeline, interpreted as a function and  $L_{\pi}$ is the left action\footref{fn2} of the rotation by $\pi$.
In the equivariant pipeline, the loss function in Eq.~\ref{eq:symloss} becomes trivial, as both $C_4$ and $D_4$ include the rotation by $\pi$, and thus $||\hat{Y}_i-\Gamma (Y_i)||=||R_{\pi}[\hat{Y}_i]-\Gamma (R_{\pi}[Y_i])||$.
When we use $H=C_4$, which includes rotations but no reflections, we adapt the symmetric loss above by replacing $L_{\pi}$ by $L_f$, where $f$ is the reflection in Eq.~\ref{eq:element_g}. 
If $L_f$ is used, we refer to the loss as `mirror~loss' $\mathcal{L}_{\text{mir}}$.

\section{Experiments}
\begin{wrapfigure}[23]{r}{0.45\textwidth}  
  \centering
  \vspace*{-2.3em}
  \begin{subfigure}[]{\linewidth}
    \centering
    \includegraphics[width=\linewidth]{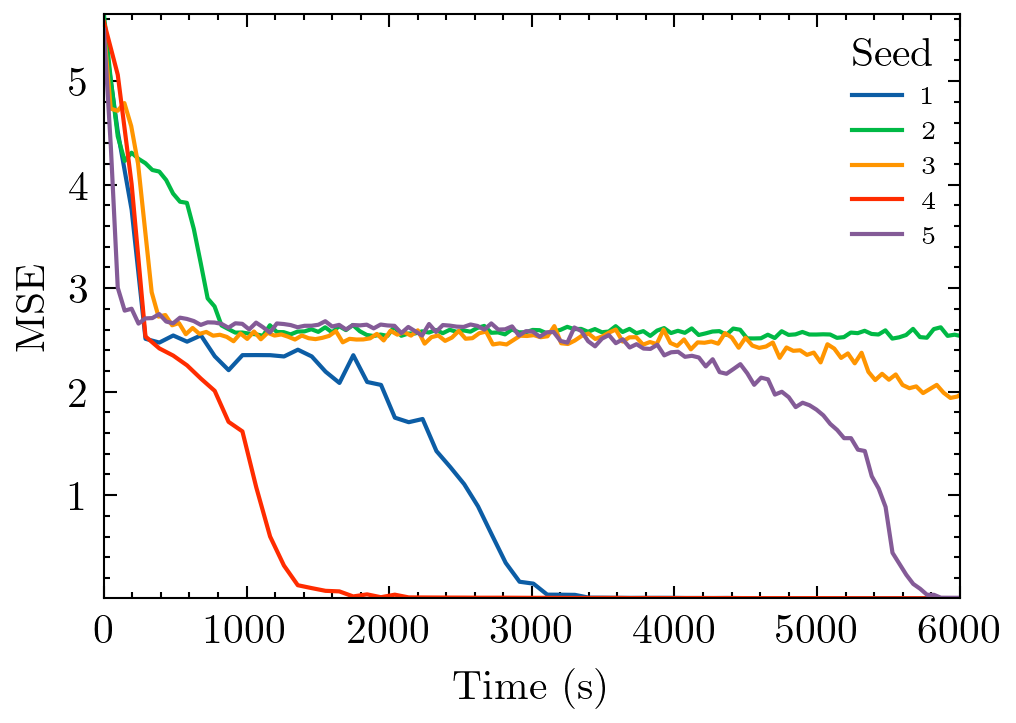}
    \label{fig:sub1}
  \end{subfigure}\\
  \begin{subfigure}[]{\linewidth}
    \centering
    \vspace*{-1em}
    \includegraphics[width=\linewidth]{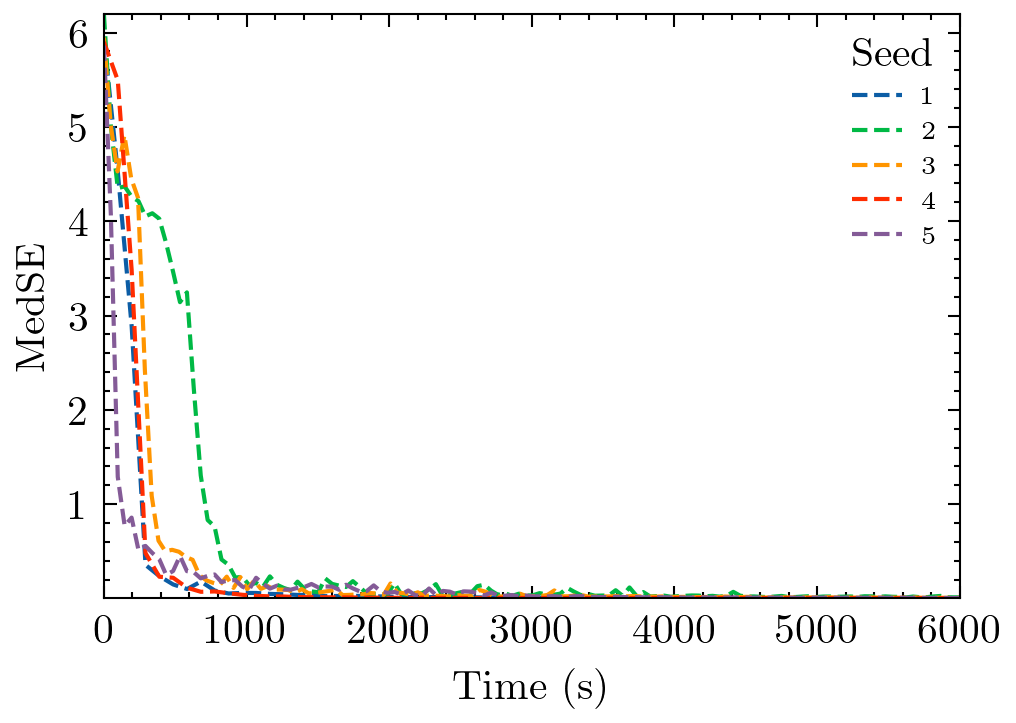}
    \label{fig:sub2}
  \end{subfigure}
  \vspace*{-2em}
  \caption{\small The MSE and MedSE over time for each of the five runs on the \textit{noiseless}  \texttt{spliceosome} dataset.}
  \label{fig:different_seeds_es}
\end{wrapfigure}
Simulated datasets with varying noise levels were generated as in \cite{levy2022cryoai} from the following Protein Data Bank structures: plasmodium falciparum 80S ribosome (PDB: 3J79 and 3J7A) \cite{wong2014cryo} (\texttt{80S}), and the pre-catalytic spliceosome (PDB: 5NRL) \cite{plaschka2017structure} (\texttt{spliceosome}). See Apx.~\ref{app:data} for more details. 

\subsection{Convergence of original pipeline}
\label{conv_standard_pipeline}

To examine the convergence progress of the original pipeline, cryoAI was trained with a non-equivariant encoder and $\mathcal{L}_{\text{sym}}$ across datasets for two structures - each with a noiseless and an $\text{SNR} = 1$ version - using $5$ different seeds for 6000 seconds.
Fig.~\ref{fig:different_seeds_es} illustrates the trends of Mean Squared Error (MSE) and Median Squared Error (MedSE) for $5$ runs using a noiseless \texttt{spliceosome} dataset. The observed pattern shows an initial decline in both MedSE and MSE, with MedSE approaching zero fast and MSE plateauing, then sharply decreasing. 
This behavior, further analyzed in Apx.~\ref{app:expl}, correlates with transient spurious planar symmetries in volume reconstructions, which resolve as MSE improves. 
While $\mathcal{L}_{\text{sym}}$ reduces the duration of such states, they still occur.
\looseness=-1

\subsection{Impact of equivariance on convergence of pipeline}

\begin{figure}[h!]
    \centering
    \includegraphics[scale=.9]{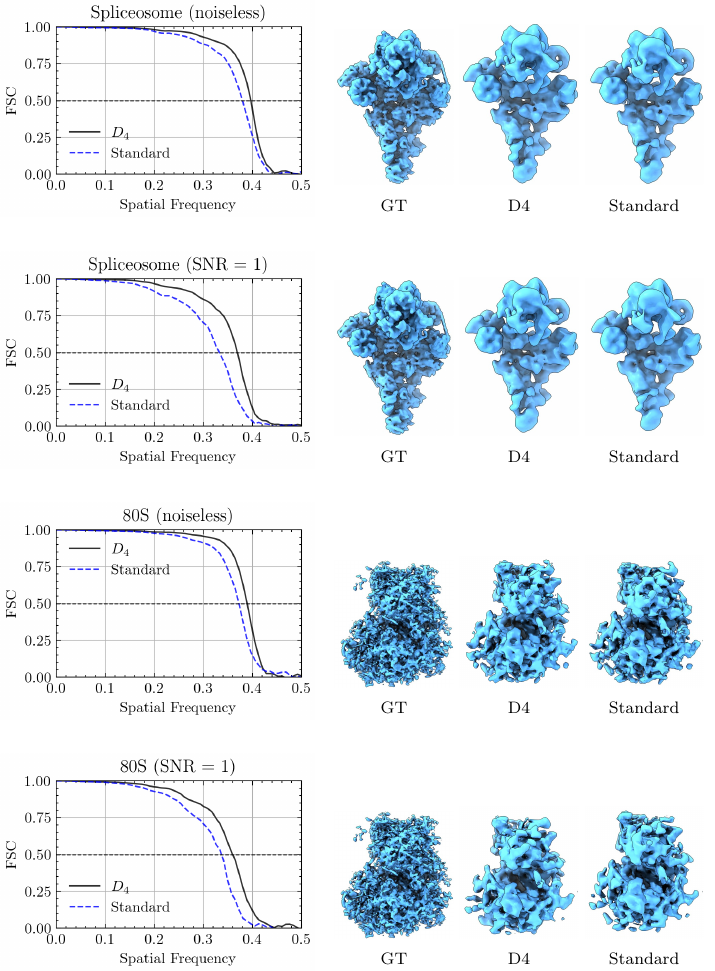}
\caption{Fourier Shell Correlation (FSC) curves and reconstruction visualizations comparing the reconstruction to the ground truth (GT) for the top-performing seeds in 6000-second runs across four simulated datasets. Each dataset is represented by two curves and two visualizations: one from the best run using the $D_4$-equivariant model and one from the best run with the standard cryoAI pipeline. The volume visualisations were made with software ChimeraX (\cite{meng2023ucsf}). The $D_4$-equivariant model consistently improves the FSC score over the non-equivariant cryoAI baseline.}    
\label{fig:fsc}
\end{figure}

We evaluated $H\!=\!C_4$- and $D_4$-equivariant encoders against a standard model using \(\mathcal{L}_{\text{mir}}\) or an $L2$ loss across four simulated datasets (each for two structures, with and without noise), with $8$ seeds per dataset and a 6000-second training limit. 
Performance metrics including MSE, convergence rates, and volume resolution are summarized in Tab.~\ref{tab:main}, with metric definitions in Apx.~\ref{app:metrics}. Convergence is defined as MSE $<$ 0.1 based on experiments with the standard pipeline that showed that the volume escapes symmetric states after the MSE has decreased sufficiently.
Our findings reveal that \(C_4\) equivariance with $L2$ loss rarely converges, but it \textit{significantly enhances convergence speed and accuracy when paired with}  \(\mathcal{L}_{\text{mir}}\).
\(D_4\)-equivariant encoders \textit{reliably converge without} \(\mathcal{L}_{\text{mir}}\), outperforming all other models in speed of convergence and metric performance, as highlighted by the superior resolution and pose estimation accuracy.
Moreover, we compare both the Fourier Shell Correlation (FSC) curves and the 3D reconstructions of some of the models in Fig.~\ref{fig:fsc}. 
The $D_4$-equivariant models \emph{consistently yield higher FSC curves, indicating improved reconstruction quality}.
However, we note that the improved reconstruction quality might still not be enough to provide additional biological insights.
Since we are not sufficiently expert to judge this aspect, we leave it for future exploration but emphasize the computational benefits of our method with respect to the standard cryoAI pipeline, particularly in terms of faster reconstruction.

\begin{table}[h!]
\caption{\small Average values for the MSE, MedSE, convergence time in seconds, number of iterations until convergence, resolution in Angstrom and the percentage of runs that converged. The average is taken over the three best runs in terms of final MSE. The resolution is reported using the $\text{FSC} = 0.5$ criterion. For each metric, the best value is indicated in bold. A dash (-) indicates that at least one of the three best runs did not converge.}
\small  
\setlength{\tabcolsep}{2.5pt}
\begin{tabular}{llllllll}
\hline
Dataset                                                                          & Model                                                               & MSE                             & MedSE                          & \begin{tabular}[c]{@{}l@{}}Conv. \\ time (s)\end{tabular} & \begin{tabular}[c]{@{}l@{}}Steps \\ until \\ conv.\end{tabular} & \begin{tabular}[c]{@{}l@{}}Resolution \\ (Å)\end{tabular} & \begin{tabular}[c]{@{}l@{}} Conv. \\ Runs \\ \%\end{tabular} \\ \hline
\multirow{4}{*}{\begin{tabular}[c]{@{}l@{}}\texttt{Spliceosome}\\ noiseless\end{tabular}} & Standard {\tiny ($\mathcal{L}_{\text{sym}}$)}   & 0.0040 {\tiny $\pm$ 0.0002}  & 0.0030 {\tiny $\pm$ 0.0001} & 2818 {\tiny $\pm$ 286} & 5800 {\tiny $\pm$ 589} & 11.26 {\tiny $\pm$ 0.11} & 62.5 \\
                                                                                 & $C_4$ {\tiny (L2)}                                       & 2.5893 {\tiny $\pm$ 0.0322} & 0.0166 {\tiny $\pm$ 0.0090} & - & - & 12.18 {\tiny $\pm$ 0.13} & 0 \\
                                                                                 & $C_4$ {\tiny ($\mathcal{L}_{\text{mir}}$)}               & 0.0008 {\tiny $\pm$ 0.0001} & 0.0005 {\tiny $\pm$ 0.0000} & 1457 {\tiny $\pm$ 1001} & \textbf{1967 {\tiny $\pm$ 1347}} & 11.11 {\tiny $\pm$ 0.19} & 100 \\
                                                                                 & $D_4$ {\tiny (L2)}                                       & \textbf{0.0004 {\tiny $\pm$ 0.0000}} & \textbf{0.0003 {\tiny $\pm$ 0.0000}} & \textbf{1239 {\tiny $\pm$ 782}} & 2800 {\tiny $\pm$ 1766} & \textbf{10.88 {\tiny $\pm$ 0.18}} & 87.5 \\ \hline
\multirow{4}{*}{\begin{tabular}[c]{@{}l@{}}\texttt{80S}\\ noiseless\end{tabular}}         & Standard {\tiny ($\mathcal{L}_{\text{sym}}$)}   & 0.0041 {\tiny $\pm$ 0.0001}  & 0.0030 {\tiny $\pm$ 0.0001} & 2564 {\tiny $\pm$ 394} & 5300 {\tiny $\pm$ 804} & 10.20 {\tiny $\pm$ 0.10} & 100 \\
                                                                                 & $C_4$ {\tiny (L2)}                                       & 0.8160 {\tiny $\pm$ 1.1520}  & 0.0020 {\tiny $\pm$ 0.0016} & - & - & 10.35 {\tiny $\pm$ 0.28} & 25 \\
                                                                                 & $C_4$ {\tiny ($\mathcal{L}_{\text{mir}}$)}               & 0.0010 {\tiny $\pm$ 0.0000} & 0.0006 {\tiny $\pm$ 0.0000} & 791 {\tiny $\pm$ 195} & \textbf{1067 {\tiny $\pm$ 262}} & 10.05 {\tiny $\pm$ 0.00} & 100 \\
                                                                                 & $D_4$ {\tiny (L2)}                                       & \textbf{0.0005 {\tiny $\pm$ 0.0001}} & \textbf{0.0004 {\tiny $\pm$ 0.0000}} & \textbf{774 {\tiny $\pm$ 143}} & 1767 {\tiny $\pm$ 330} & \textbf{9.72 {\tiny $\pm$ 0.09}} & 87.5 \\ \hline
\multirow{4}{*}{\begin{tabular}[c]{@{}l@{}}\texttt{Spliceosome}\\ SNR 1\end{tabular}}     & Standard {\tiny ($\mathcal{L}_{\text{sym}}$)}   & 0.3232 {\tiny $\pm$ 0.3384}  & 0.0077 {\tiny $\pm$ 0.0031} & - & - & 12.72 {\tiny $\pm$ 0.97} & 12.5 \\
                                                                                 & $C_4$ {\tiny (L2)}                                       & 2.6525 {\tiny $\pm$ 0.0104}  & 0.0308 {\tiny $\pm$ 0.0273} & - & - & 13.06 {\tiny $\pm$ 0.15} & 0 \\
                                                                                 & $C_4$ {\tiny ($\mathcal{L}_{\text{mir}}$)}               & 0.0014 {\tiny $\pm$ 0.0001} & 0.0010 {\tiny $\pm$ 0.0001} & \textbf{1181 {\tiny $\pm$ 312}} & \textbf{1600 {\tiny $\pm$ 432}} & 11.33 {\tiny $\pm$ 0.00} & 87.5 \\
                                                                                 & $D_4$ {\tiny (L2)}                                       & \textbf{0.0010 {\tiny $\pm$ 0.0000}} & \textbf{0.0007 {\tiny $\pm$ 0.0000}} & 1624 {\tiny $\pm$ 426} & 3333 {\tiny $\pm$ 998} & \textbf{11.03 {\tiny $\pm$ 0.10}} & 50 \\ \hline
\multirow{4}{*}{\begin{tabular}[c]{@{}l@{}}\texttt{80S}\\ SNR 1\end{tabular}}             & Standard {\tiny ($\mathcal{L}_{\text{sym}}$)}   & 0.0102 {\tiny $\pm$ 0.0003}  & 0.0070 {\tiny $\pm$ 0.0009} & 3093 {\tiny $\pm$ 418} & 6400 {\tiny $\pm$ 909} & 11.22 {\tiny $\pm$ 0.00} & 75 \\
                                                                                 & $C_4$ {\tiny (L2)}                                       & 0.0055 {\tiny $\pm$ 0.0014} & 0.0037 {\tiny $\pm$ 0.0009} & 4080 {\tiny $\pm$ 999} & 10100 {\tiny $\pm$ 2491} & 10.97 {\tiny $\pm$ 0.20} & 37.5 \\
                                                                                 & $C_4$ {\tiny ($\mathcal{L}_{\text{mir}}$)}               & 0.0038 {\tiny $\pm$ 0.0002} & 0.0027 {\tiny $\pm$ 0.0001} & 1386 {\tiny $\pm$ 808} & \textbf{1867 {\tiny $\pm$ 1087}} & 10.97 {\tiny $\pm$ 0.00} & 100 \\
                                                                                 & $D_4$ {\tiny (L2)}                                       & \textbf{0.0032 {\tiny $\pm$ 0.0001}} & \textbf{0.0023 {\tiny $\pm$ 0.0000}} & \textbf{1071 {\tiny $\pm$ 135}} & 2433 {\tiny $\pm$ 309} & \textbf{10.65 {\tiny $\pm$ 0.22}} & 100 \\ \hline
\end{tabular}
\label{tab:main}
\end{table}

\paragraph{Higher noise levels}
To investigate if the convergence benefits of equivariance are maintained under higher noise conditions, we conducted an experiment focusing on the standard model and the $D_4$-equivariant model. These models were trained on datasets generated using the \texttt{80S} ribosome structure at various noise levels; see Apx.~\ref{app:data}.
Each model is trained for 6 hours. 
We recorded the metrics for the best run out of $5$ for each model and dataset.
The results in Tab.~\ref{noisytable} show that the benefits of equivariance hold at higher noise levels. 
The equivariant encoder outperforms the standard encoder on all metrics for each noise level, except the highest ($\text{SNR}=0.25$), although it should be noted that all seeds of both models didn't converge at that noise level.
\begin{table}[b]
\centering
\caption{\small Pose estimation error, convergence time, and final resolution of the best model across 5 seeds for both model types for each of the noise levels. Resolution is reported using the~$\text{FSC} = 0.5$~criterion.}
\small
\begin{tabular}{llllll}
\hline
SNR                                       & Model                               & MSE        & MedSE      & \begin{tabular}[c]{@{}l@{}}Convergence time (s)\end{tabular} & \begin{tabular}[c]{@{}l@{}}Final resolution (Å)\end{tabular} \\ \hline
\multicolumn{1}{c}{\multirow{2}{*}{0.75}} & $D_4$ {\tiny (L2)}             & \textbf{0.0012}     & \textbf{0.0009}     & \textbf{1238.04}                                                         & \textbf{9.46}             \\
\multicolumn{1}{c}{}                      & Standard {\tiny ($\mathcal{L}_{\text{sym}}$)}          & 0.0027     & 0.0021     & 3065.58                                                         & 9.65             \\ \hline
\multirow{2}{*}{0.625}                    & $D_4$ {\tiny (L2)}             & \textbf{0.0013}     & \textbf{0.0011}     & \textbf{1136.72}                                                         & \textbf{9.56}             \\
                                          & Standard {\tiny ($\mathcal{L}_{\text{sym}}$)}          & 0.0036     & 0.0025     & 2722.84                                                         & 9.85             \\ \hline
\multirow{2}{*}{0.5}                      & $D_4$ {\tiny (L2)}             & \textbf{0.0021}     & \textbf{0.0014}     & \textbf{1850.60}                                                         & \textbf{9.75}             \\
                                          & Standard {\tiny ($\mathcal{L}_{\text{sym}}$)}          & 0.0057     & 0.0039     & 3885.72                                                         & 10.72            \\ \hline
\multirow{2}{*}{0.375}                    & $D_4$ {\tiny (L2)}             & \textbf{0.0057}     & \textbf{0.0033}     & \textbf{6263.10}                                                         & \textbf{10.97}            \\
                                          & Standard {\tiny ($\mathcal{L}_{\text{sym}}$)}          & 0.0780     & 0.0096     & 12741.46                                                        & 12.37            \\ \hline
\multirow{2}{*}{0.25}                     & $D_4$ {\tiny (L2)}             & 1.4671     & 0.1627     & -                                                               & 22.98            \\
                                          & Standard {\tiny ($\mathcal{L}_{\text{sym}}$)}          & \textbf{0.7455}     & \textbf{0.0546}     & -                                                               & \textbf{16.09}            \\ \hline
\end{tabular}
\label{noisytable}
\end{table}

\section{Final Discussions}

Our experiments demonstrate the benefits of incorporating geometric priors in the Cryo-EM amortized inference framework.
Indeed, equivariant models consistently show drastic improvements in convergence speed, convergence ratio and final resolution.
In particular, $D_4$-equivariance makes the expensive symmetric loss superfluous.
These results indicate that apart from the symmetric loss, equivariant amortized inference can be a viable alternative solution to the convergence issues associated with symmetric reconstructions. 
Still, some runs with equivariant encoder did not convergence;
in this context, \cite{esmaeili2023topological} suggested 
that, {while inductive biases like equivariance can guide models towards learning homeomorphic representations, they might also complicate the training process in practice. 
Inspired by their solution, we find encoders modeling a multimodal distribution, as opposed to our single-pose encoders, to be a promising direction of research.}
\looseness=-1


\newpage

\appendix

\section{Image formation model in Fourier space}
\label{app:im_fourier}

Implementing the image formation model in reconstruction algorithms can become computationally expensive due to the rotation of the volume $V$ and the integral in equation \ref{eq:P_i}. Formulating the image formation in Fourier space allows for the application of the \textit{Fourier Slice Theorem}, which makes rotating and integrating the volume redundant.

The theorem states that evaluating a volume on a plane (`a slice of coordinates') in Fourier space gives the Fourier transform of a projection of that volume along an axis perpendicular to that plane. In mathematical terms, this means that
\begin{equation}
   S_R[ \mathcal{F}_{3}(V)]= \mathcal{F}_2(P_R[V])
   \label{eq:slice}
\end{equation}
where $\mathcal{F}_2$ and $\mathcal{F}_3$ are the two- and three dimensional Fourier transforms, and $S_R$ is an operator that performs the operation of evaluating a volume on a plane, and is defined such that
\begin{equation}
    S_{R}[V]\colon (k_1,k_2) \to V(R^{\top}\cdot (k_1,k_2,0)^{\top}).
\end{equation}
\noindent Using the \textit{Fourier Slice Theorem}, we can formulate the entire image formation model in Fourier space as follows:
\begin{equation}
    \hat{I}_i=\hat{h}_i \odot \hat{T}_{t_i} \odot S_{R_i}[\mathcal{F}_3(V)] + \hat{\epsilon}_i
    \label{eq:image_formation_fourier}
\end{equation}
Here $\hat{h}_i$ is the Fourier transform of the $h_i$ kernel, called the Contrast Transfer Function (CTF), $\hat{T}_{t_i}$ is the translation operator performing $t_i$ in Fourier space (which corresponds to a phase shift) and $\hat{\epsilon}_i$ is i.i.d. complex Gaussian noise at each frequency. The symbol $\odot$ stands for element-wise multiplication that replaces convolution in the real formulation, which follows from the Convolution Theorem.

\section{Decoder architecture}
\label{app:decoder}
The decoder takes the rotation $R_i$ and the translation $t_i$ outputted by the encoder as an input. An array of $L^2$ three-dimensional coordinates $[k_x,k_y,0]^{\top}\in \mathbb{R}^3$ is generated that represent the Fourier frequencies corresponding to a grid of $L^2$ coordinates on the $xy-$plane. Each coordinate in this slice is then rotated by $R_i^{\top}$ and fed into a neural network $\hat{V}\colon \mathbb{R}^3\to \mathbb{C}$. The network $\hat{V}$, called FourierNet, represents the Fourier transform of the electrostatic potential of the molecule that we wish to reconstruct. 

\noindent FourierNet consists of two SIREN models that take a coordinate $[k_x,k_y,k_z]\in \mathbb{R
}^3$ and output a vector in $\mathbb{R}^2$. SIREN (Sinusoidal Representation) networks are fully connected feed-forward neural networks that utilize sinusoidal activation functions. SIREN networks have been shown to be successful in representing complex signals defined on real space. (\cite{sitzmann2020implicit}) The Fourier transform of an electrostatic of a molecule is known to vary in magnitude over several orders of magnitude. (\cite{levy2022cryoai}) To allow FourierNet to represent a function that varies several orders of magnitude in value, one of the two SIRENs has the property that the exponential function is applied element-wise on its output. 

\noindent To ensure that the network $\hat{V}$ actually represents a Fourier transform, the following property is enforced
\begin{equation}
    \hat{V}(k)=\hat{V}(-k)^* \,\,\,\,\,\,\,\,\,\,\,\text{if } k_x<0
\end{equation}
simply by following this as a definition. The two output vectors of the networks are multiplied element-wise and the result is interpreted as an element of $\mathbb{C}$ via the mapping
\[
[v_1, v_2]^{\top}\mapsto v_1+v_2 i.
\]
Since the input to FourierNet was a slice of three-dimensional coordinates, the output represents a slice of a Fourier transform $S_{R_i}[\mathcal{F}_3(V)]\in \mathbb{C}^{L\times L}$ as defined in Eq. $\ref{eq:slice}$. The final output $X_i$ of the decoder is obtained as follows:
\begin{equation}
    \hat{X}_i=\hat{h}_i \odot \hat{T}_{t_i} \odot S_{R_i}[\mathcal{F}_3(V)],
\end{equation}
where $\hat{h}_i$ is the CTF, and $\hat{T}_{t_i}$ is the translation operator performing the translation $t_i$ predicted by the encoder in Fourier space. The CTF is a function of parameters that can be estimated from the dataset and are assumed to be known. For more details about the pipeline architecture, see \cite{levy2022cryoai}.

\section{Data}
\label{app:data}
The data was synthesized with structures from the Protein Data Bank. The two structures that were used were the plasmodium falciparum 80S ribosome structure (PDB: 3J79 and 3J7A) \cite{wong2014cryo} and the pre-catalytic spliceosome structure (PDB: 5NRL) \cite{plaschka2017structure}.

\begin{wrapfigure}[17]{r}{0.55\textwidth}  
  \centering
  \captionsetup[subfigure]{justification=centering}
  \subcaptionbox*{The plasmodium falciparum 80S ribosome volume.}[.45\linewidth]{%
    \centering 
    \includegraphics[width=0.9\linewidth]{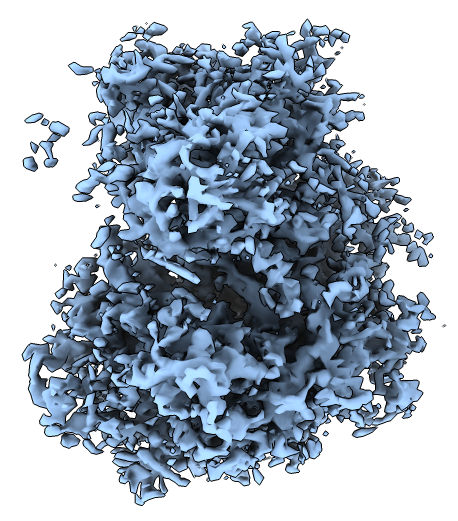}}%
  \subcaptionbox*{The pre-catalytic spliceosome volume.}[.4\linewidth]{%
    \centering 
    \includegraphics[width=0.9\linewidth]{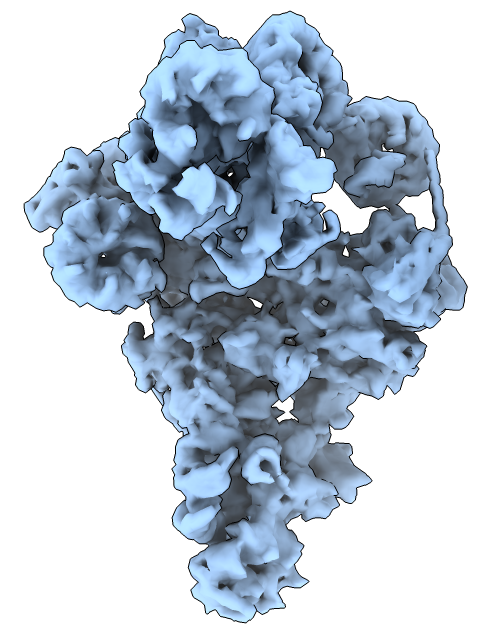}}
  \caption{Visualization of the two volumes that were used to generate the datasets. The visualisations were made with software ChimeraX \cite{meng2023ucsf}}
  \label{gtvols}
\end{wrapfigure}
The datasets were generated by simulating the image formation model described in Sec.~\ref{mathcryo} using ground-truth volume represented as a three-dimensional voxel grid, as in \cite{levy2022cryoai}. The volume data for the ground truth volumes was generated in ChimeraX \cite{meng2023ucsf} as described in Apx.~A.1 of \cite{levy2022cryoai}. See Fig.~\ref{gtvols} for a visualization of the volumes. 

Each dataset contains $100\,000$ images of dimension $128\times 128$. The simulator uses poses $R_i\in \text{SO}(3)$ randomly sampled from a uniform distribution over SO$(3)$, and translations $t_i\in \mathbb{R}^2$ sampled from a Gaussian distribution with mean zero and a standard deviation of $20 $\AA. 

\begin{figure}[h!]
  \centering
  \subcaptionbox*{Noiseless}[.14\linewidth]{%
    \centering 
    \includegraphics[width=0.9\linewidth]{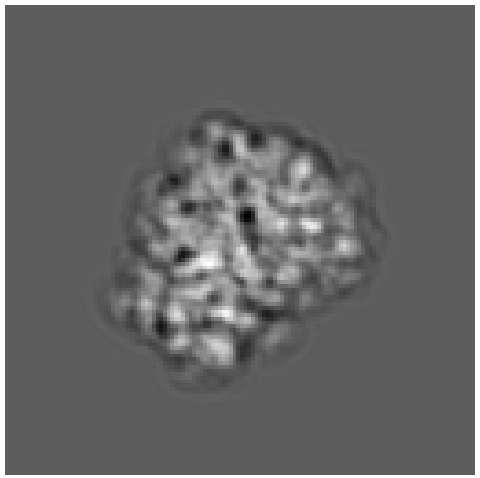}%
  }%
  \subcaptionbox*{SNR $\,1$}[.14\linewidth]{%
    \centering 
    \includegraphics[width=0.9\linewidth]{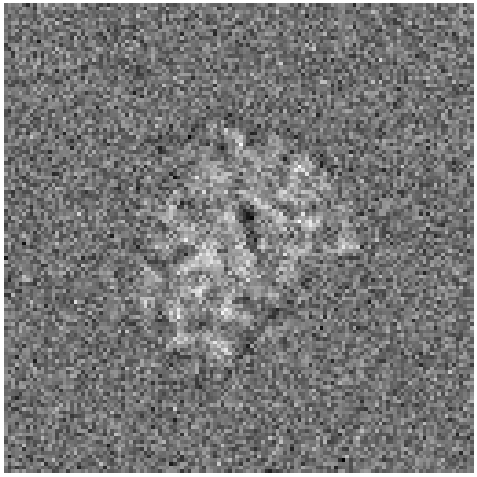}%
  }%
  \subcaptionbox*{SNR $\,0.75$}[.14\linewidth]{%
    \centering 
    \includegraphics[width=0.9\linewidth]{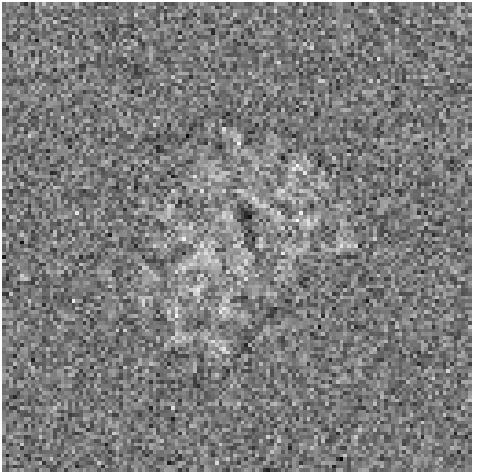}%
  }%
  \subcaptionbox*{SNR $\,0.625$}[.14\linewidth]{%
    \centering 
    \includegraphics[width=0.9\linewidth]{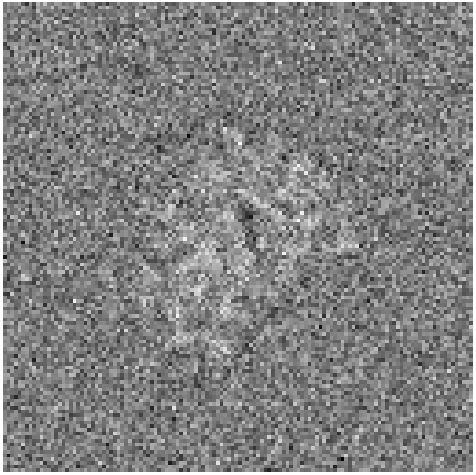}%
  }%
  \subcaptionbox*{SNR $\,0.5$}[.14\linewidth]{%
    \centering 
    \includegraphics[width=0.9\linewidth]{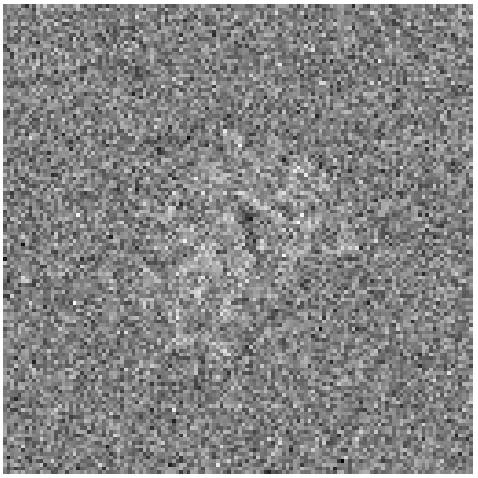}%
  }%
  \subcaptionbox*{SNR $\,0.375$}[.14\linewidth]{%
    \centering 
    \includegraphics[width=0.9\linewidth]{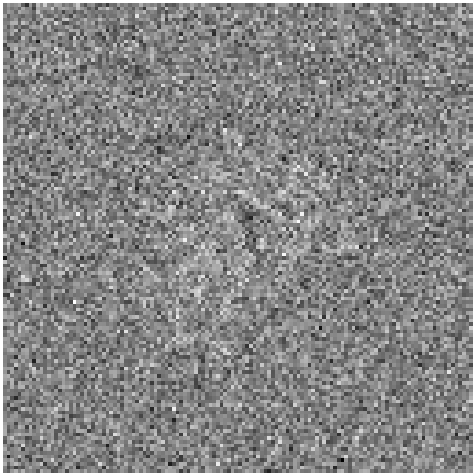}%
  }%
  \subcaptionbox*{SNR $\,0.25$}[.14\linewidth]{%
    \centering 
    \includegraphics[width=0.9\linewidth]{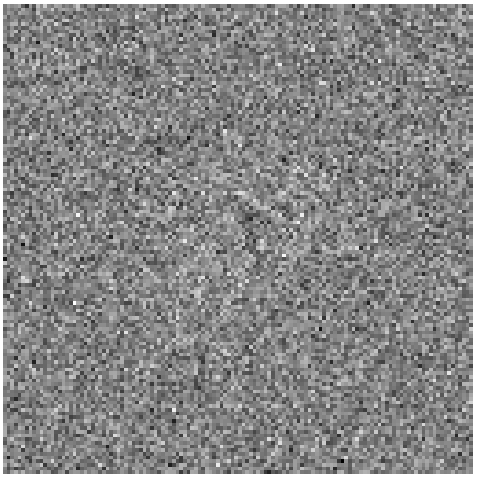}%
  }%
  \caption{Examples of images from our datasets generated with the plasmodium falciparum 80S ribosome volume with different noise levels.}
  \label{snrs}
\end{figure}

For both the pre-catalytic spliceosome and the plasmodium falciparum 80S ribosome structures, one noisless dataset was generated, and one with signal-to-noise ratio (SNR) equal to $1$. For the plasmodium falciparum 80S ribosome structure, additional datasets were generated with SNR $0.75,0.625,0.5,0.375$ and $0.25$, in Fig.~\ref{snrs} we show examples of images from each of the datasets generated with the plasmodium falciparum 80S ribosome.

\section{Background on Group and Group Representation Theory}
\label{app:group_theory}
Group theory, and specifically group representation theory, are important building blocks in the theory about equivariant neural networks, such as steerable networks that are applied in this work. In this Section we provide some important definitions and preliminaries that are useful for understanding the method presented in this work.

Groups can be useful to describe the structure of sets of geometrical transformations, such as rotations and translations. We start by giving the definition of a group.
\begin{definition}
    \paragraph{Definition} A \emph{group} is a set $G$ equipped with a group product $\cdot$ satisfying the following properties:
    \begin{itemize}
        \item Closure: If $g,h\in G$, then also $g\cdot h\in G$.
        \item Associativity: For $g,h,i \in G$, $(g\cdot h)\cdot i = g\cdot (h\cdot i)$.
        \item Identity: There exists an element $e\in G$ such that $e\cdot g=g\cdot e=g$ for all $g\in G$.
        \item Inverse element: for all $g\in G$, there exists an element $g^{-1}\in G$ such that $g\cdot g^{-1}=g^{-1}\cdot g=e$.
    \end{itemize}
\end{definition}
\begin{example}
    \paragraph{Example}
    The \emph{circle group} is the set of all elements of $\mathbb{C}$ with absolute value equal to one, denoted
    \[
    S^1=\{z\in\mathbb{C}\,|\, |z|=1\}
    \] 
    with as group product multiplication of complex numbers and as identity element 1.
\end{example}

Another important definition in group theory is that of a \emph{group homomorphism}. Intuitively, group homomorphisms are functions between groups that preserve algebraic structure shared by these groups.
\begin{definition}
    \paragraph{Definition} If $G_1$ and $G_2$ are groups, then a function $f\colon G_1\to G_2$ is a group homomorphism if for all $g,h\in G_1$
    \[
    f(g\cdot h)=f(g)\star f(h).
    \]
    Here $\cdot$ is used to denote the group product of $G_1$ and $\star$ for the group product of $G_2$. \\ \\ 
    \noindent If $G_1=G_2$, $f$ is called an automorphism. If $f$ is bijective, it is called a \emph{group isomorphism}. If there is a group isomorphism between $G_1$ and $G_2$, we call $G_1$ and $G_2$ isomorphic and we write $G_1 \cong G_2$. \\ \\
    \noindent The set of all automorphisms of a group $G$ is again a group with map composition as a group product. It is denoted by Aut$(G)$.
\end{definition}
It can be useful to `combine' several groups into one. In this way we can represent different types of transformations (e.g. translation and rotation) as elements of a single group. One way to do this is via \emph{semi-direct products}.

\begin{definition}
\paragraph{Definition} Let $H$ and $N$ be groups. Let $\tau: H\to \text{Aut}(N)$  be a group homomorphism. We define the semi-direct product of $H$ with $N$ with respect to $\tau$, written $N\rtimes_{\tau}H$, as the set $N\times H$ with the following group operation
\[
(n_1, h_1)\cdot (n_2, h_2)=(n_1\tau(h_1)(n_2), h_1h_2).
\]
\end{definition}

As mentioned, some groups describe the structure of sets of geometrical transformations, such as rotations and translations. Such groups have \emph{actions} on other spaces. Group representations are one way to describe the actions of certain groups (linear transformations) on vector spaces. To define group representations, we first give the definition of the general linear group:
\begin{definition}
    \paragraph{Definition} The general linear group of a vector space $V$, denoted GL($V$), is the set of all invertible linear maps from $V$ to $V$ with as a group product composition of maps.
\end{definition}
\noindent Since in deep learning we are often working with real-valued vectors, when we refer to the general linear group, we are usually talking about GL($\mathbb{R}^n$). This is the group of all invertible $n\times n$ matrices.

Using the definition of the general linear group, we can define group representations.
\begin{definition}
    \paragraph{Definition} A \emph{representation} of a group $G$ is a homomorphism $\varphi\colon G\to \text{GL}(V)$ for some vector space $V$. We can write $\varphi(g)$ or $\varphi_g$ for $\varphi$ evaluated at $g$.\\
    \noindent We write $d_{\varphi}$ to denote the \emph{dimension} of the representation, which is equal to the dimension of the vector space $V$.
\end{definition}

\noindent Some examples of representations follow below.

\begin{example}
    \paragraph{Example} For any group $G$ we can define the representation $\varphi\colon G\to \text{GL}(\mathbb{R})$ as $\varphi(g)=1$ for all $g\in G$. This is called the \emph{trivial representation}. 
\end{example}

\noindent By looking at a representation of the circle group $S^1$ that we defined before, we see that the circle group is actually also a group that describes a specific type of transformation on $\mathbb{R}^2$: rotations.
\begin{example}
    \paragraph{Example} We can define a representation $\varphi\colon S^1\to \text{GL}(\mathbb{R}^2)$ as follows:
    \[
    \varphi(e^{i \theta})= \begin{bmatrix}
\cos(\theta) & -\sin(\theta) \\
\sin(\theta) & \cos(\theta) 
\end{bmatrix}.
    \]
    Every element of $S^1$ is mapped onto a rotation matrix by this representation, which shows us that $S^1$ has an action on $\mathbb{R}^2$ corresponding to planar rotation around the origin. \\
    
    \noindent The group formed by all $2\times 2$ rotation matrices is called SO$(2)$.
    One can verify that $\varphi$ is a isomorphism between $S^1$ and SO$(2)$, which means that $S^1\cong \text{SO}(2)$.
\end{example}
We can combine multiple representations together into a single representation via the \emph{direct sum}. This concept is important in the theory of steerable CNNs.
\begin{definition}
    \paragraph{Definition} Let $\rho_1\colon G\to \text{GL}(V_1)$ and $\rho_2\colon G\to \text{GL}(V_2)$ be two representations of $G$. Then $(\rho_1 \oplus \rho_2)\colon G\to \text{GL}(V_1\oplus V_2)$, given by
    \begin{equation}
        (\rho_1 \oplus \rho_2)(g) = \begin{bmatrix}\rho_1(g) & 0 \\ 0 & \rho_2(g)\end{bmatrix}
    \end{equation}
    is called their direct sum. The direct sum $(\rho_1 \oplus \rho_2)$ is again a representation. Since it is a diagonal matrix with $\rho_1$ and $\rho_2$ on the diagonals, it works on the vector space $(V_1\oplus V_2)$ by letting $\rho_1$ work on the subspace $V_1$, and $\rho_2$ on the subspace $V_2$.
\end{definition}

An important representation in the theory of steerable CNNs is the \emph{regular representation}. 
\begin{example}
    \paragraph{Example} Let $G$ be a finite group. If $f\colon G\to \mathbb{R}$ is a function over $G$ and $g\in G$, then
    \[
    [g.f](h)=f(g^ {-1}h)
    \]
    describes an action of $G$ on the vector space of functions over $G$. The regular representation $\rho_{\text{reg}}$ of $G$ is the representation that describes this action. \\
    
    \noindent The vector space of functions over $G$ is equivalent to $\mathbb{R}^{|G|}$, since we can interpret a function $f\colon G\to \mathbb{R}$ as a vector where the $i$th element represents the function value assigned to the $i$th group element $g_i$. For an element $g\in G$ and $f\in \mathbb{R}^{|G|}$, $\rho_{\text{reg}}$ associates to $g$ the permutation matrix that sends element $i$ to the element $j$ such that $g^{-1}\cdot g_i=g_j$.
\end{example}

We finish this section with some important groups that we regularly refer to in this work.
The first group, SO$(3)$, contains all three-dimensional rotation matrices.
\begin{example}
\paragraph{Example} The \emph{special orthogonal group} \( \text{SO}(3) \) consists of all \( 3 \times 3 \) real orthogonal matrices \( R \) with a determinant of \( +1 \), representing three-dimensional rotation matrices. Specifically, 
\[
\text{SO}(3) = \{ R \in \mathbb{R}^{3 \times 3} \mid R^TR = I \text{ and } \det(R) = 1 \}.
\]
Here, \( R^T \) denotes the transpose of \( R \), \( I \) is the \( 3 \times 3 \) identity matrix, and the group operation is matrix multiplication.
\end{example}
Another important group, $O(2)$, contains two-dimensional rotations and reflections.
\begin{example}
\paragraph{Example} The \emph{orthogonal group} \( O(2) \) consists of all \( 2 \times 2 \) real orthogonal matrices \( A \), which includes both rotations and reflections in two-dimensional space. It is defined as:
\[
O(2) = \{ A \in \mathbb{R}^{2 \times 2} \mid A^TA = I \},
\]
where \( A^T \) is the transpose of \( A \), and \( I \) is the \( 2 \times 2 \) identity matrix. The group operation is matrix multiplication.
\end{example}
Finally, the groups $C_4$ and $D_4$ are both subgroups of O$(2)$.
\begin{example}
\label{c4}
\paragraph{Example} The \emph{cyclic group} \( C_4 \) consists of the set of rotational symmetries of a square, containing four elements corresponding to rotations by multiples of 90 degrees. It is defined as:
\[
C_4 = \{ e, r, r^2, r^3 \},
\]
where \( e \) is the identity element (no rotation), and \( r \) is a 90-degree rotation. The group operation is the composition of rotations.
\end{example}

\begin{example}
\paragraph{Example} \label{d4}
The \emph{dihedral group} \( D_4 \), representing the symmetries of a square, consists of eight elements: four rotations and four reflections. It can be described as:
\[
D_4 = \{r, s \mid r^4 = s^2 = 1, \, srs = r^{-1} \},
\]
here \( r \) represents a 90-degree rotation, \( s \) represents a reflection, and \( r^{-1} \) is the inverse of \( r \).
\end{example}

\section{Equivariant encoder architecture}
\label{app:eq_enc_arch}
\begin{figure}
    \centering
    \includegraphics[scale=1]{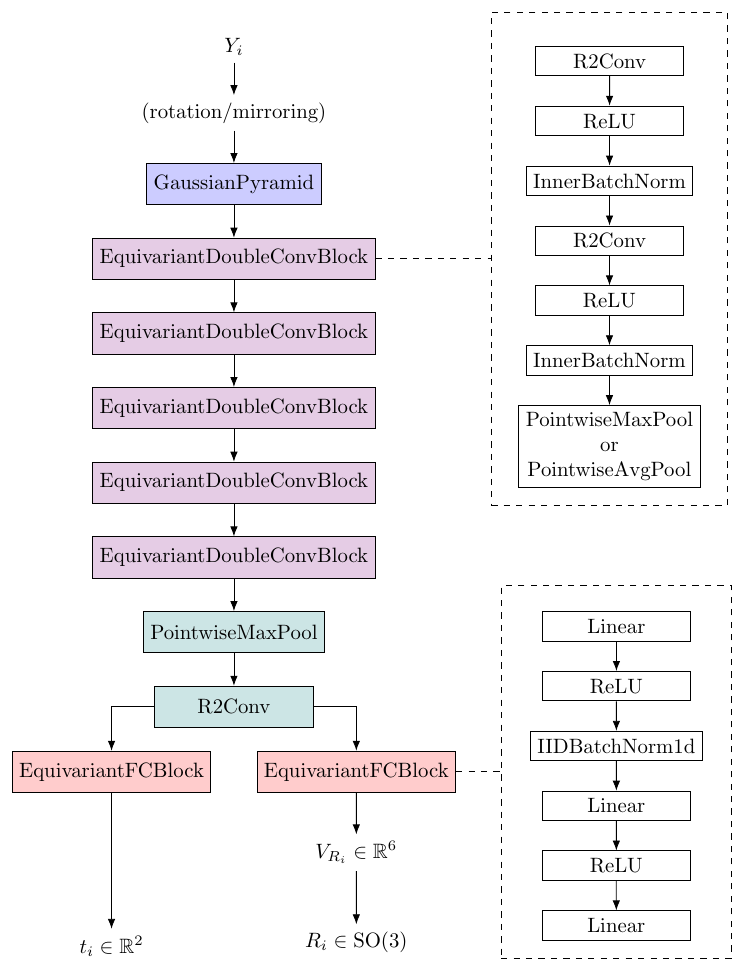}
    \caption{Visualization of our architecture of the equivariant encoder.}
    \label{fig:eq_encoder}
\end{figure}

A high-level overview of the equivariant encoder is shown in Figure \ref{fig:eq_encoder}. As mentioned in Section \ref{pipeline}, the architecture was kept as similar as possible to the standard pipeline in \cite{levy2022cryoai}. The encoders were implemented as steerable CNNs \cite{cohen2016steerable} with the \texttt{escnn} Python library \cite{weiler2019general, cesa2022a}

If a symmetric- or mirror loss function (see Section \ref{sec:loss}) is used, the input $Y_i$ is fed into a sym layer. This layer performs the operation of duplicating the batch of images and then rotating it by $\pi$ or mirroring it with respect to the $x$-axis (depending on the type of loss). The transformed images are concatenated batch-wise. If an L$2$ loss function is used, this process is skipped. The batch then goes into a GaussianPyramid layer with five Gaussian filters of size $11$ with cutoff frequencies that are geometrically distributed between 0.1 and 10 pix$^{-1}$ (as in \cite{levy2022cryoai}), that filter each image. 

The filtered images are fed into five consecutive {EquivariantDoubleConvBlock} networks. Each {EquivariantDoubleConvBlock} layer is made up of two convolutional layers with steerable kernels (R2Conv) and a PointwiseMaxPool layer. Each of the R2Conv layers is followed by a pointwise ReLU activation function and an InnerBatchNorm layer. 

After passing through each of the EquivariantDoubleConvBlock blocks, the batch passes through a PointwiseMaxPool layer and a single R2Conv layer with a kernel of size $2$. This results in an output with a width and height of $1$, with overal shape
$
\text{(batch size)} \times \text{(number of feature fields}\cdot |G|)\times \text{(1)} \times \text{(1)}.
$
This output is fed into two separate equivariant fully connected networks (EquivariantFCBlock). The first EquivariantFCBlock outputs a vector $t_i\in\mathbb{R}^2$, which is directly interpreted as a translation. The second EquivariantFCBlock outputs a vector $V_{R_i}\in\mathbb{R}^6$, which is then mapped to $R_i\in \text{SO}(3)$ and interpreted as a rotation matrix as explained in 
Apx.~\ref{app:encoder_details}.

\subsection{Technical details}
\label{app:encoder_details}
In this section we present some technical details related to the implementation of the encoder. We start with some very brief theory.

Recall that in regular CNNs, layers are convolution operators that, when applied to an input, yield a \emph{feature map}. A key difference between steerable CNNs and regular CNNs is that in the former, feature maps are replaced by steerable feature fields:

\begin{definition}
\paragraph{Definition} A steerable feature field is a vector field $f:\mathbb{R}^2 \to \mathbb{R}^c$ defined on the base space $\mathbb{R}^2$, together with a transformation law that defines how the group $G=\mathbb{R}^2 \rtimes H$ acts on the field $f$ as a whole.

The transformation law given an element $g=(t, h)\in G$ has the following form:
\begin{equation}
    [g.f](x) = \rho(h) f(g^{-1}x).
    \label{eq:transf_law}
\end{equation}
where $\rho: H \to \text{GL}(\mathbb{R}^c)$ is a representation of the group $H$. It defines how $H$ acts on $\mathbb{R}^c$. The representation $\rho$ is called the \emph{type} of this particular feature field.
\end{definition} 

Layers in steerable CNNs are equivariant, meaning that if a feature field is transformed by a group element according to its transformation law and then passes through a layer, we want this to give the same output as the feature field first passing through that layer and then transforming the generated feature field by the same group elements according to its own transformation law. Mathematically, we can write this as:
\begin{equation}
    (k\star t_l(g)[ f]) = t_{l+1}(g) [(k\star f)]
    \label{eq:equivariance}
\end{equation}
for all $g\in G$. Here $t_n$ is the transformation law of the feature field in layer $n$ (defined as in equation \ref{eq:transf_law}), and $\star$ is the convolution operator.

\paragraph{Implementation} As mentioned, the equivariant encoder was implemented using steerable CNNs. As part of the implementation of networks with the \texttt{escnn} library, one specifies both an \emph{input type} and an \emph{output type} for each layer, which together define the type of equivariance of the layer exhibits (as explained above).

The input type of the first DoubleConv block is a direct sum of five trivial representations, since its input is a stack of five filtered images that transform separately. The input type of each consecutive DoubleConv block is a direct sum of multiple regular representations. The number of regular representations in the direct sum is a hyperparameter that is different for each of the encoders. The features in each layer in the EquivariantFCBlock also transform according to a direct sum of multiple regular representations. 

The output types of both EquivariantFCBlocks are especially important, since they determine the overall type of equivariance the encoder exhibits. Recall that we want the encoder to satisfy the following equations:
\begin{equation}
    E_r(L_g[I]) = \varphi_r(g)E_r(I),
    \label{eq:rot_eq2}
\end{equation}
and
\begin{equation}
    E_t(L_g[I]) = \varphi_t(g)\cdot E_t(I).
    \label{eq:tr_eq2}
\end{equation}
where $\varphi_t(g):G\to \text{GL}(\mathbb{R}^2)$ and $\varphi_r(g):G\to \text{GL}(\mathbb{R}^3)$ are two representations of $G$ such that for $g= f^c\cdot r_{\theta}\in G$,
\begin{equation}\varphi_t(g)=\begin{bmatrix}1 & 0 \\0 & -1\end{bmatrix}^c\begin{bmatrix} \cos(\theta) & -\sin(\theta) \\ \sin(\theta) & \cos(\theta) \end{bmatrix}\end{equation}
and
\begin{equation}\varphi_r(g)=\begin{bmatrix}
    -1 & 0 & 0 \\
    0 & 1 & 0 \\
    0 & 0 & -1
    \end{bmatrix}^c \begin{bmatrix}
    \cos(\theta) & -\sin(\theta) & 0 \\
    \sin(\theta) & \cos(\theta) & 0 \\
    0 & 0 & 1
    \end{bmatrix}\end{equation}

$E_t$ satisfies equation \ref{eq:tr_eq2} if the output type $\rho^{\text{out}}_{E_t}$ of the first EquivariantFCBlock is equal to $\varphi_t$, since $E_t$ directly outputs $t_i$. 

For the second EquivariantFCBlock we choose output type $\rho^{\text{out}}_{E_r}=\varphi_r \oplus \varphi_r$. Recall that the output of this EquivariantFCBlock is a vector $V_{R_i}\in\mathbb{R}^6$, which is then mapped to $R_i\in \text{SO}(3)$ and interpreted as a rotation matrix (see Figure \ref{fig:eq_encoder}). Therefore, we also have to define a mapping $V_{R_i}\mapsto R_i$ in such a way that equation \ref{eq:rot_eq2} is satisfied. This is done as follows:

Let \( V_{R_i} = [v_1\,\, \dots\,\, v_6]^{\top} \), \( V_1 = [v_1\,\, v_2\,\, v_3]^{\top} \), \( V_2 = [v_4\,\, v_5\,\, v_6]^{\top} \) and \( V_3 = V_1 \times V_2 \), where $\times$ indicates the cross product\footnote{The cross product \( V_3 = V_1 \times V_2 \) is given by $[v_2v_6 - v_3v_5 \,\,\,v_3v_4 - v_1v_6\,\,\,v_1v_5 - v_2v_4]^\top$. The direction of $V_3$ is perpendicular to the plane formed by $V_1$ and $V_2$, according to the right-hand rule}. We define
\begin{equation} 
M = \begin{bmatrix}
 | &  | & | \\
 V_1 & V_2 & V_3 \\
 | &  | & |
\end{bmatrix}
\end{equation}
Let $U_M,S_M,V_M$ be the Singular Value Decomposition (SVD) of \( M \), and define \( D_M \) as the diagonal matrix where \( D_{ii} = 1 \) for \( i = 1, 2 \) and \( D_{33} = \det(UV^{\top}) \). Then we define a mapping $O:M\mapsto R_i$ such that it performs the \emph{orthogonalization} of the matrix $M$:
\begin{equation}
    R_i = U_MD_MV_M^{\top}
\end{equation}
The orthogonalisation of a matrix $M$ is an operation that projects it to the nearest element in SO$(3)$ in the least-squares sense (\cite{levinson2020analysis}).

The cross-product has the following general property. If \( R \) is an orthogonal matrix in $\mathbb{R}^{3\times 3}$, then we have the identity:
\begin{equation}
\label{identitycross}
    (RV_1) \times (RV_2)=R(V_1 \times V_2) 
\end{equation} 
Note that since $V_{R_i}$ transforms according to $\rho^{\text{out}}_{E_r}=\varphi_r \oplus \varphi_r$, a block-diagonal matrix where the blocks are two copies of the three-dimensional representation $\varphi_r$, $V_1$ and $V_2$, and according to the identity in equation \ref{identitycross} also $V_3$ transform according to $\varphi_r$ ($\varphi_r$ is orthogonal). Since $V_1,V_2$ and $V_3$ are the columns of $M$, all the columns of the matrix $M$ transform according to $\varphi_r$.

To prove that equation \ref{eq:rot_eq2} is satisfied, we need to prove that the orthogonalization mapping $O:M\mapsto R_i$ is equivariant, that is
\begin{equation}
\label{ortho}
    O(\varphi_r M)=\varphi_r O(M).
\end{equation}
We can prove this by using the previously mentioned property of orthogonalization of a matrix that it is an operation that projects a matrix to the nearest element in SO$(3)$ in the least-squares sense:
\begin{equation}
    O(M)=\argmin_{R\in \text{SO}(3)} ||R-M||^2_F
\end{equation}
We assume that the singular values of $M$ and are distinct, so that we can assume that the singular value decomposition is unique. Then we can prove equation \ref{ortho} via a proof by contradiction as follows:

Suppose that $O(M)=\argmin_{R\in \text{SO}(3)} ||R-M||^2_F=R$ and suppose that equation \ref{ortho} does not hold, that is $O(\varphi_r M)=R'\neq \varphi_r R$. Recall that we assumed that the singular value decomposition was unique, thus we can assume that $R'$ and $R$ are the unique elements of SO(3) minimizing the norm in equation \ref{ortho}. Using that the Frobenius norm is invariant under multiplication with orthogonal matrices, we can derive
\begin{align}
    ||R'-\varphi M||_F^2& < ||\varphi R-\varphi M||_F^2\\
    ||\varphi_r^{\top}(R'-\varphi M)||_F^2& < ||\varphi_r^{\top}(\varphi R-\varphi M)||_F^2\\
    ||\varphi^{\top}R'-\varphi^{\top}\varphi M||_F^2& < ||\varphi^{\top}\varphi R-\varphi^{\top}\varphi M||_F^2\\
    ||\varphi^{\top}R'- M||_F^2& < ||R- M||_F^2.
\end{align}
This is a contradiction, since we assumed that $R=\argmin_{R\in \text{SO}(3)} ||R-M||^2_F$. So we must have that equation \ref{ortho} holds. 

\section{Metrics}
\label{app:metrics}
\paragraph{Pose prediction error} The reconstructed volume and therefore also the rotations predicted by the encoder given images in the dataset differ from their ground truth (used for the simulation of the data) by some global rotation $R^{\text{G}}\in \text{SO}(3)$. Therefore, before determining the pose prediction error we first have to find an estimation for $R^{\text{G}}$ and use it to re-align the predicted poses to the ground truth poses. We follow a method based on publicly available code provided by the authors of \cite{levy2022cryoai}. In this method, the mean squared error ($\text{MSE}$) and median squared error (MedSE) for a given dataset $\mathcal{D}=\{I_1,\dots,I_N\}$ of $N$ images are determined as follows. For every $I_i\in \mathcal{D}$, we write $R^{\text{pred}}_i := E_r(I_i)$ for the rotation predicted by the encoder, and $R^{\text{gt}}_i$ for the rotation that was used in the simulation of the dataset to generate $I_i$. Let REL be the set containing the relative rotations that align each of the predicted rotations to their corresponding ground-truth rotation:
\begin{equation}
    \text{REL}:=\{R^{\text{rel}}_i\,|\,\text{for all }1\leq i\leq N\text{ such that }R^{\text{rel}}_i R^{\text{pred}}_i= R^{\text{gt}}_i \}.
\end{equation}
We then choose the rotations in REL that minimize the mean squared Frobenius norm and the median squared Frobenius norm between the aligned predicted rotations and the ground truth rotations to calculate the MSE and MedSE:
\begin{equation}
\label{eq:mse}
    \text{MSE}:=\min_{R\in \text{REL}} \frac{1}{N}\sum_{i}^N \lVert RR^{\text{pred}}_i - R^{\text{gt}}_i \rVert^2_{\text{F}}, \text{ and}
\end{equation}
\begin{equation}
\label{eq:medse}
    \text{MedSE}:=\min_{R\in \text{REL}} \text{median}\left(\lVert RR^{\text{pred}}_1 - R^{\text{gt}}_1 \rVert^2_{\text{F}},\,\dots,\,\lVert RR^{\text{pred}}_N - R^{\text{gt}}_N \rVert^2_{\text{F}}\right).
\end{equation}

\paragraph{Resolution} The Fourier Shell Correlation (FSC) is a widely utilized method in cryo-EM to assess the resolution of 3D reconstructions. It quantifies the degree of correlation between two independent reconstructions obtained from different sets of data.

Let \( V_1 \) and \( V_2 \) represent two 3D reconstructed volumes in real space. The Fourier transforms of these volumes are denoted as \( \hat{V}_1\) and \( \hat{V}_2 \). The FSC is computed as a function of the spatial frequency, and for a given shell in the Fourier space (defined by a specific frequency magnitude \( |k| \)), it is given by:
\begin{equation}
\text{FSC}(|k|) = \frac{\sum_{|k|} \hat{V}_1(k) \cdot \overline{\hat{V}_2(k)}}{\sqrt{\sum_{|k|} |\hat{V}_1(k)|^2 \cdot \sum_{|k|} |\hat{V}_2(k)|^2}}
\end{equation}
where \( \overline{\hat{V}_2(k)} \) denotes the complex conjugate of \( \hat{V}_2(k) \).

The FSC curve, plotted as a function of spatial frequency, provides insight into the consistency of structural information between the two reconstructions at different resolutions. A higher FSC value indicates a higher degree of correlation, suggesting better agreement between the reconstructions. The resolution at which the FSC curve drops to a threshold value (commonly set to 0.143 or 0.5) is typically taken as an indicator of the resolution of the reconstructed volume. In this work, we determine the resolution by comparing the reconstructed volume with the corresponding ground truth volume used to create the simulated data. We set the threshold to 0.5 to determine the resolution of the volume.
 
\section{Exploration of convergence process of standard pipeline}
\label{app:expl}
As described in Section \ref{conv_standard_pipeline}, the standard cryoAI pipeline optimized with a symmetric loss function shows a convergence pattern in which there is an initial decline in both MedSE and MSE, with MedSE approaching zero fast and MSE plateauing, then sharply decreasing (Figure \ref{fig:different_seeds_es}). 

Given how we define the MSE and the MedSE (see Section \ref{app:metrics}, Equation \ref{eq:mse} and Equation \ref{eq:medse}), this suggests that a subset of at least half of the predicted poses can be aligned to the ground truth with a low error relatively quickly, while the the other subset of poses show a higher error with this alignment. 

\paragraph{Internally consistent subsets} Consider a subset of images \( S \subseteq \mathcal{D}\) of the full dataset $\mathcal{D}$. For any image \( I_i \) in \( S \) and its corresponding predicted rotation \( R^{\text{pred}}_i \), we define the relative rotation \( R^{\text{rel}}_i \) such that \( R^{\text{rel}}_i R^{\text{pred}}_i = R^{\text{gt}}_i \). This rotation aligns the predicted rotations with the ground truth rotation. If the encoder of the model is consistent on the subset $S$, we expect \( R^{\text{rel}}_i \) to align the predicted rotations of all images in \( S \) with their ground truths (\( R^{\text{rel}}_i R^{\text{pred}}_j = R^{\text{gt}}_j \) for all \( j \) where \( I_j \) is in \( S \)). Therefore, if an image \( I_i \) is part of an internally consistent subset \( S_i \), counting the number of images \( I_j \) where \( R^{\text{rel}}_i R^{\text{pred}}_j \approx R^{\text{gt}}_j \) gives an approximate size of \( S_i \). In mathematical terms, for a given $i$, we can define $S_i$ as follows:
\begin{equation}
    S_i = \{I_j : || R_i^{rel} R_j^{pred} - R_j^{gt} || < \epsilon\} \subseteq D.
\end{equation}

If the encoder makes prediction for all images in a dataset \( \mathcal{D} \) with low error, each subset \( S_i \) would be equal to the entire dataset (\( |S_i| = |\mathcal{D}| \) for every \( I_i \) in \( \mathcal{D} \)). Conversely, if the encoder is in a state where it classifies half of the images into one internally consistent subset \( S \) and the other half into another subset \( S' \), then \( |S| = |S'| = \frac{1}{2}|\mathcal{D}| \). We found that in states where the MSE was high whereas the MedSE had already decreased, the latter was the case.

\begin{figure}[h!]  
  \centering
 \includegraphics[scale=0.5]{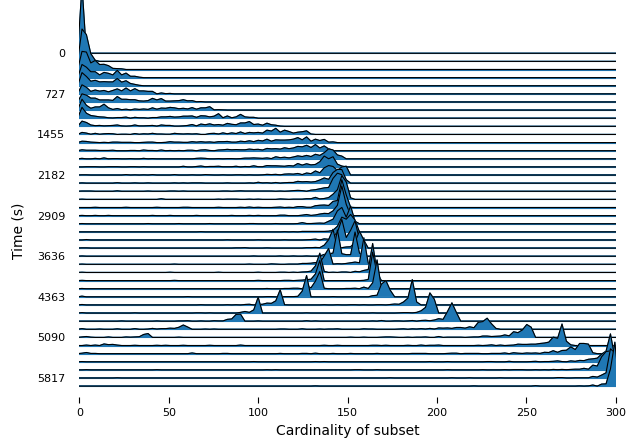}
  \caption{Joy plot with on each horizontal line the distribution of the cardinalities of subsets $|S_i|$ for the first 300 images in our dataset $\mathcal{D}$ at a given time step.}
  \label{fig:histograms}
\end{figure}
We show this in Figure \ref{fig:histograms}, where we illustrate the distribution of \( |S_i| \) for the first 300 images in the training dataset \( \mathcal{D} \), observed over time in a run exhibiting the characteristic trajectory mentioned earlier. We notice that after some time, every image belongs to a group of size 150, which is half of the data points. This indicates the existence of two internally consistent subsets of images. This state persists for a while, after which one subset starts shrinking while the other grows until it encompasses all data points. The moment one subset begins to shrink and the other to grow coincides with the decrease in MSE.

\paragraph{Projective plane visualization} To explore the convergence process of the encoder, we developed a way to visualize the way in which the encoder predicts poses on the projective plane. 

Let \( \mathcal{D} = \{I_1, I_2, \dots, I_n\} \) represent the set of images from a cryo-EM experiment. We define the equivalence relation \( \sim \) on \( \mathcal{D} \) such that \( I_i \sim I_j \) if and only if there exists some \( g \in \text{O}(2) \) where \( I_j = L_g[I_i] \) (where $L_g$ is the left action of $g$, see footnote \ref{fn2}). This relation leads to the formation of the set of equivalence classes \( \mathcal{D}/{\sim} \).

Since each image \( I_i \) in \( \mathcal{D} \) corresponds to an element \( R_i \) in SO(3), each element in \( \mathcal{D}/{\sim} \) can be seen as part of the real projective plane P\(\mathbb{R}^2 \cong \) SO(3)/O(2). The projective plane can be imagined as the unit sphere $S^2$ where each pair of opposite points is identified. Recall that each image has a corresponding pose \( R_i = (x_i, y_i, z_i) \) in SO(3), with \( z_i \) being the axis of projection in the image formation process, also known as the `viewing direction'. Each equivalence class \( [I_i] \) in \( \mathcal{D}/{\sim} \) then corresponds to a point on the projective plane, where the viewing direction \( z_i \) is one of such paired points.

\begin{wrapfigure}[19]{r}{0.5\textwidth}  
    \centering
      \begin{tikzpicture}[scale=0.7]
    \def\r{2}
    
    \pgfmathsetmacro{\zcoord}{sqrt((\r*\r)-0.5)-0.2}
    
    \draw[thick] (0.5,\zcoord,-0.5) -- (0,0,0);
    \draw[dotted,thick] (0.5,\zcoord,-0.5) -- (0.5,0,-0.5);
    \draw[thick] (-0.5,-\zcoord,0.5) -- (0,0,0);
    
    \shade[ball color = gray!40, opacity = 0.4] (0,0,0) circle (\r);

    \begin{scope}[canvas is zx plane at y=0]
    \draw[->] (0,0,0) -- (0,1.5*\r,0) node[anchor=north west]{$x$}; 
    \draw[->] (0,0,0) -- (-1.5*\r,0,0) node[anchor=north]{$y$}; 
    \draw[color=black,opacity=0.3] (1.5*-\r,1.5*-\r,0) -- (1.5*\r,1.5*-\r,0) -- (1.5*\r,1.5*\r,0) -- (1.5*-\r,1.5*\r,0) -- cycle;

    \pgfplotsset{view={130}{30}}
   \end{scope}
    \draw[->] (0,0,0) -- (0,1.5*\r,0) node[anchor=south]{$z$}; 
    
    \node at (0.5,\zcoord,-0.5) [circle,fill,inner sep=1pt]{};
    \node at (0.5,0,-0.5) [circle,fill,inner sep=1pt]{};
    \node at (-0.5,-\zcoord,0.5) [circle,fill,inner sep=1pt]{};
    \node at (-0.5,-\zcoord,0.5) [left] {$z_i$};
\end{tikzpicture}
    \caption{We interpret a pose $R_i = (x_i, y_i, z_i)$ as a point on the projective plane and visualize this as depicted. If $z_i=(z_{i1},z_{i2},z_{i3})$, then we project $R_i$ to the point $(\text{sgn}(z_{i3})z_{i1},\text{sgn}(z_{i3})z_{i1})$ on the plane. Here sgn is the function that returns the sign of its input.}
    \label{fig:projplane}
\end{wrapfigure}
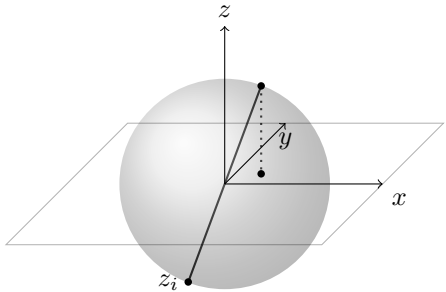

We make the projective plane visualizations as shown in Figure \ref{fig:projplane}, by projecting poses onto the xy-plane as shown \emph{after first aligning them with the ground truth poses} in the same way as we would to to determine the MSE (see \ref{app:metrics}). We do this for both the ground truth poses and the poses predicted by the encoder. To visualize how the predicted poses differ from their ground truth pose, we color each of the predicted poses with the color that their corresponding ground truth pose has according to a colormap on the xy-plane.

\begin{figure}[h!]
  \centering
    \captionsetup[subfigure]{justification=centering}
  \subcaptionbox*{Ground truth poses}[0.3\linewidth]{%
    \centering 
    \includegraphics[width=0.9\linewidth]{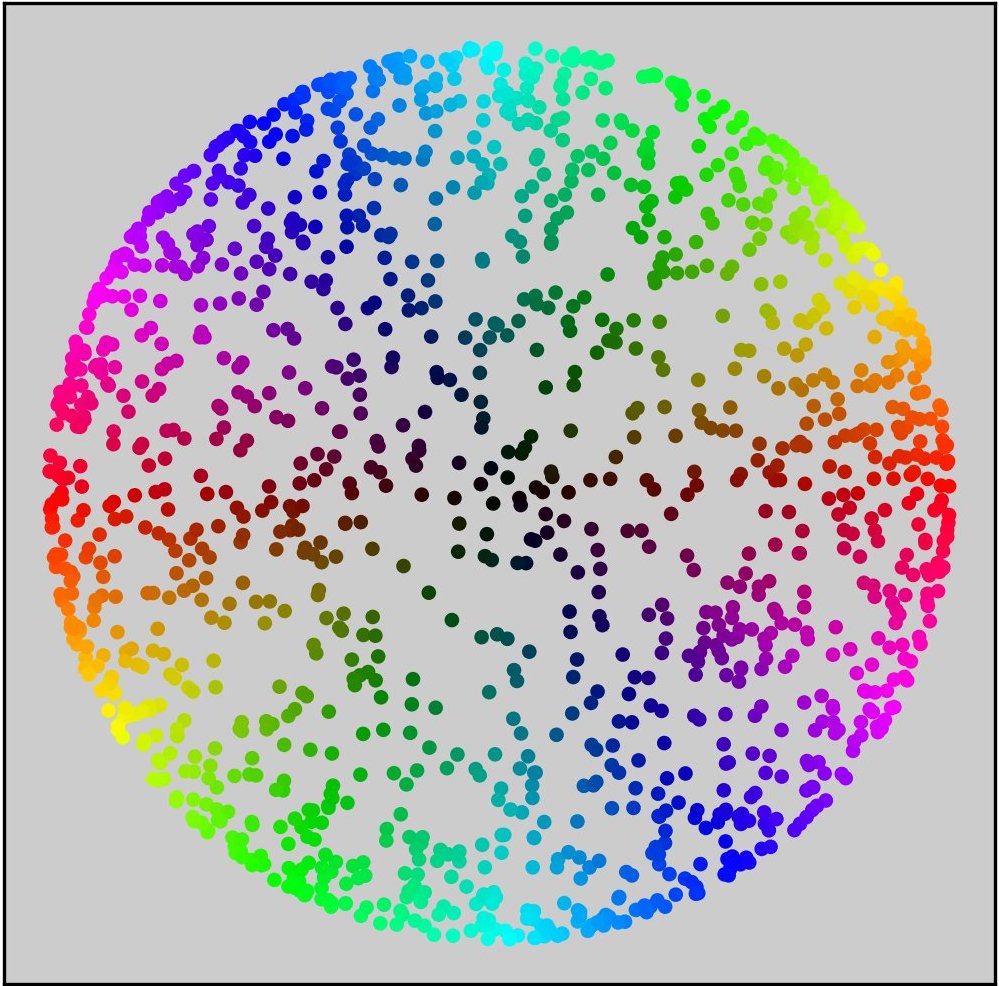}%
  }%
  \subcaptionbox*{Pattern 1 of predicted poses}[0.3\linewidth]{%
    \centering 
    \includegraphics[width=0.9\linewidth]{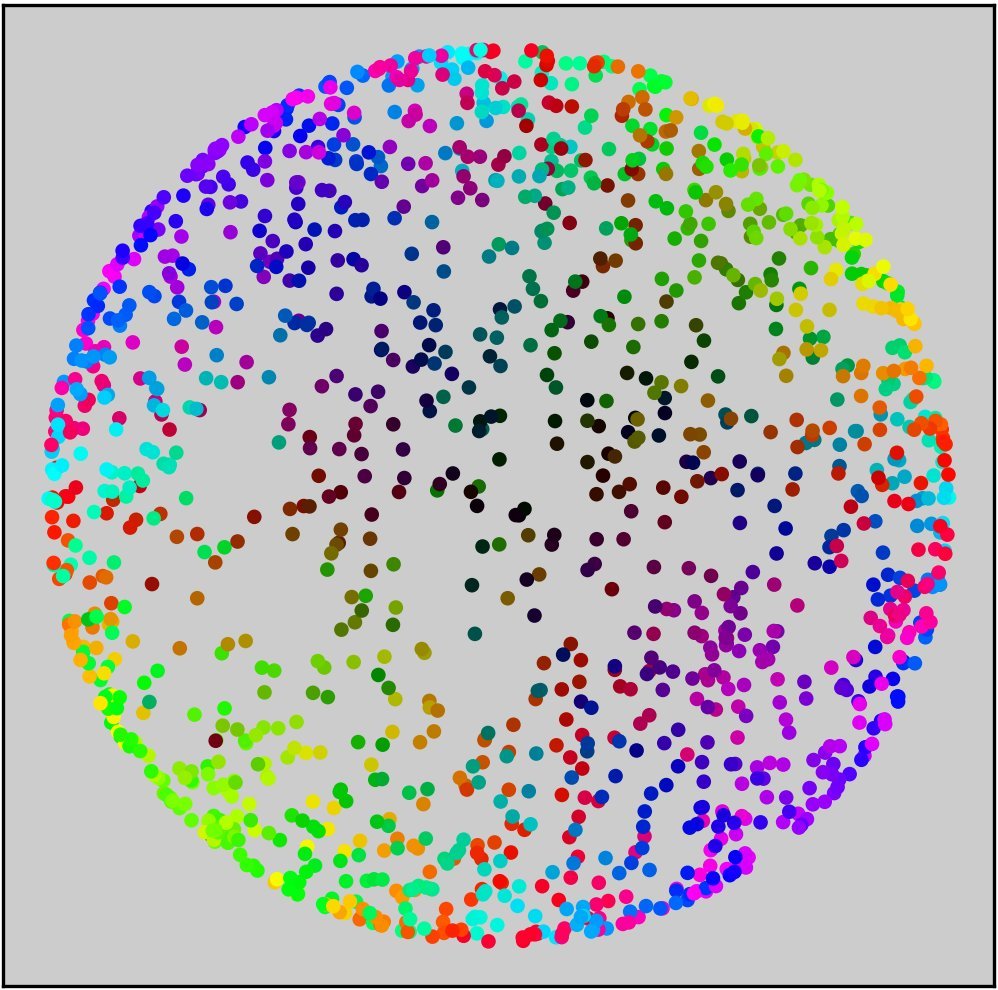}%
  }
    \subcaptionbox*{Pattern 2 of predicted poses}[0.3\linewidth]{%
    \centering 
    \includegraphics[width=0.9\linewidth]{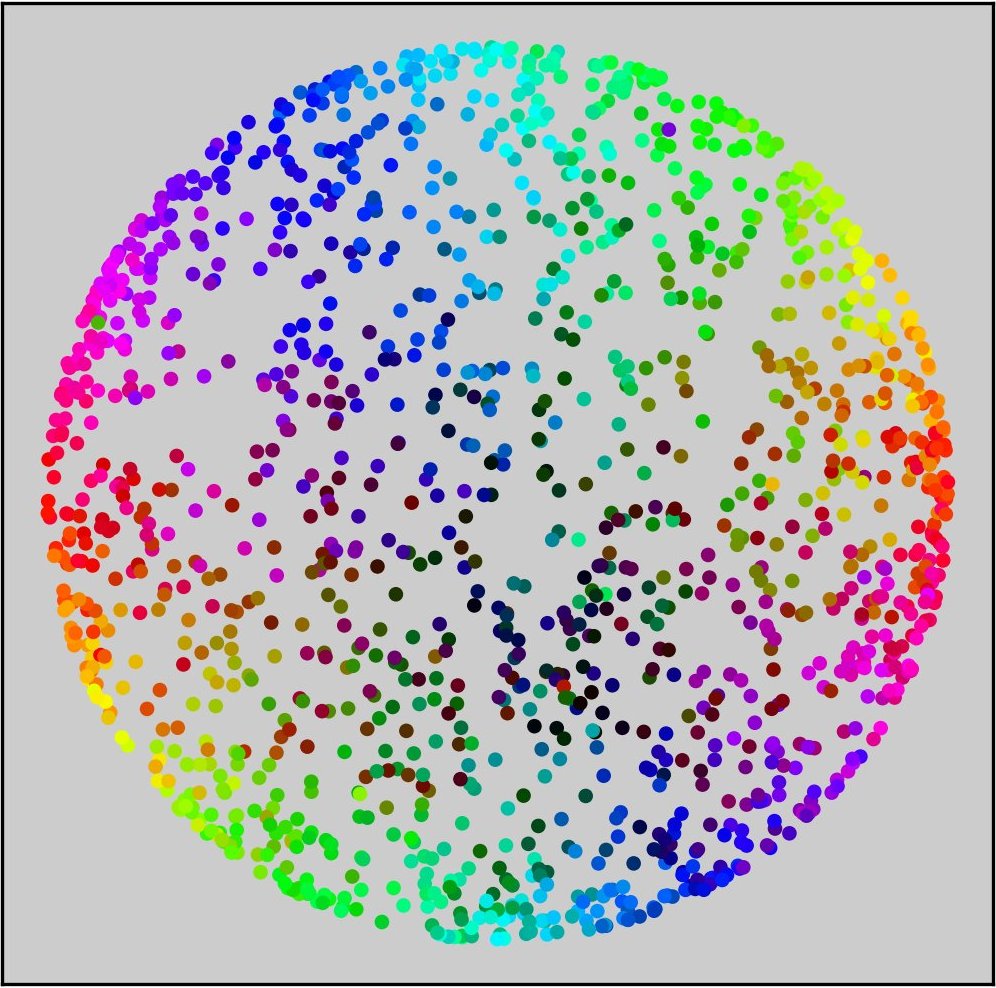}%
  }
  \caption{Patterns observed in projective plane visualisations for runs trained on datasets generated with the 80S ribosome structure.}
  \label{pp80s}
\end{figure}

\begin{figure}[h!]
  \centering
    \captionsetup[subfigure]{justification=centering}

  \subcaptionbox*{Ground truth poses}[0.3\linewidth]{%
    \centering 
    \includegraphics[width=0.9\linewidth]{Figures/gt.jpg}%
  }%
  \subcaptionbox*{Pattern of predicted poses}[0.3\linewidth]{%
    \centering 
    \includegraphics[width=0.9\linewidth]{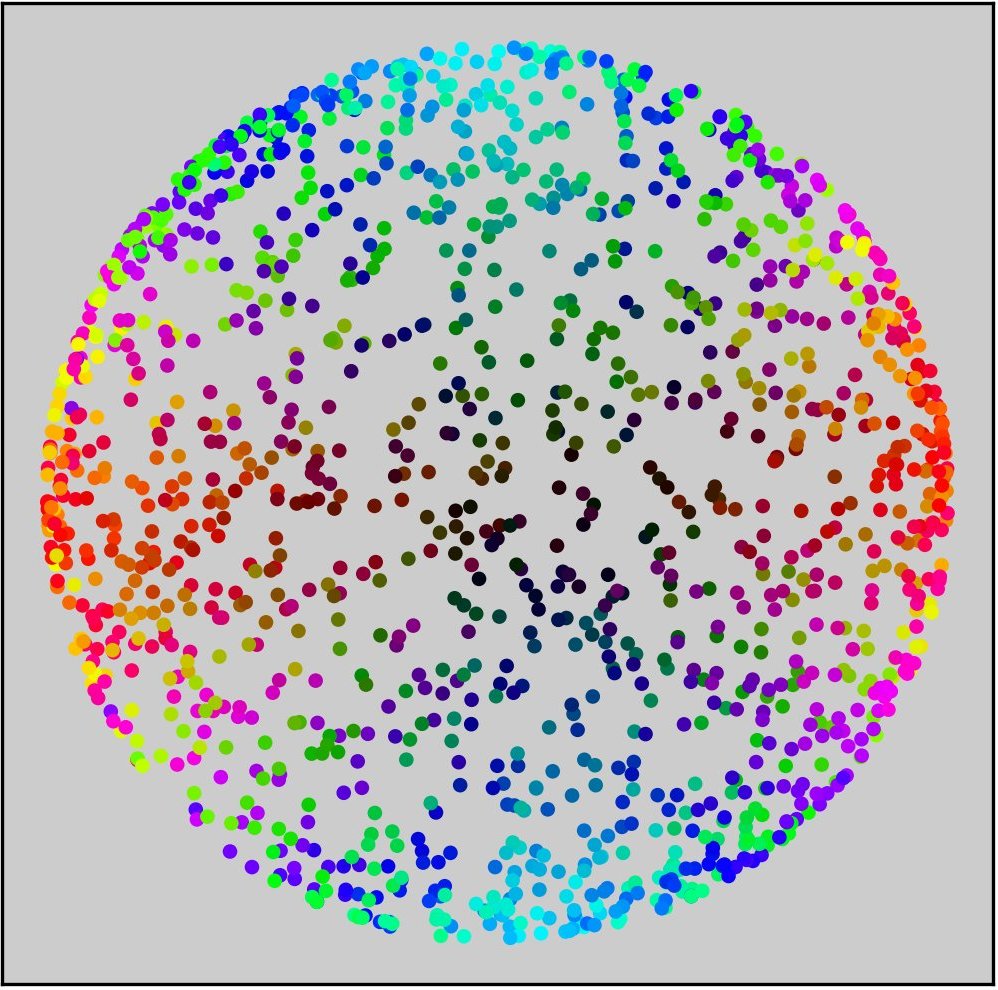}%
  }
  \caption{Patterns observed in projective plane visualisations for runs trained on datasets generated with the spliceosome structure.}
  \label{ppspl}
\end{figure}

In Figure \ref{ppspl} and Figure \ref{pp80s} we show examples of patterns in the projective plane visualizations that occur for the different datasets in states where the MSE is high, and the MedSE is low. For both structures, a pattern is observable where approximately half of the point are mirrored with respect to some line in the xy-plane. From these patterns in the projective plane alone, interpreting how the encoder misclassifies subsets of poses is challenging. However, the patterns indicate that the internally consistent subsets of data points are sometimes distinguishable within the projective plane. 

\begin{figure}[h!]
  \centering
      \captionsetup[subfigure]{justification=centering}

  \subcaptionbox*{Spliceosome structure}[.40\linewidth]{%
    \centering 
    \includegraphics[width=0.8\linewidth]{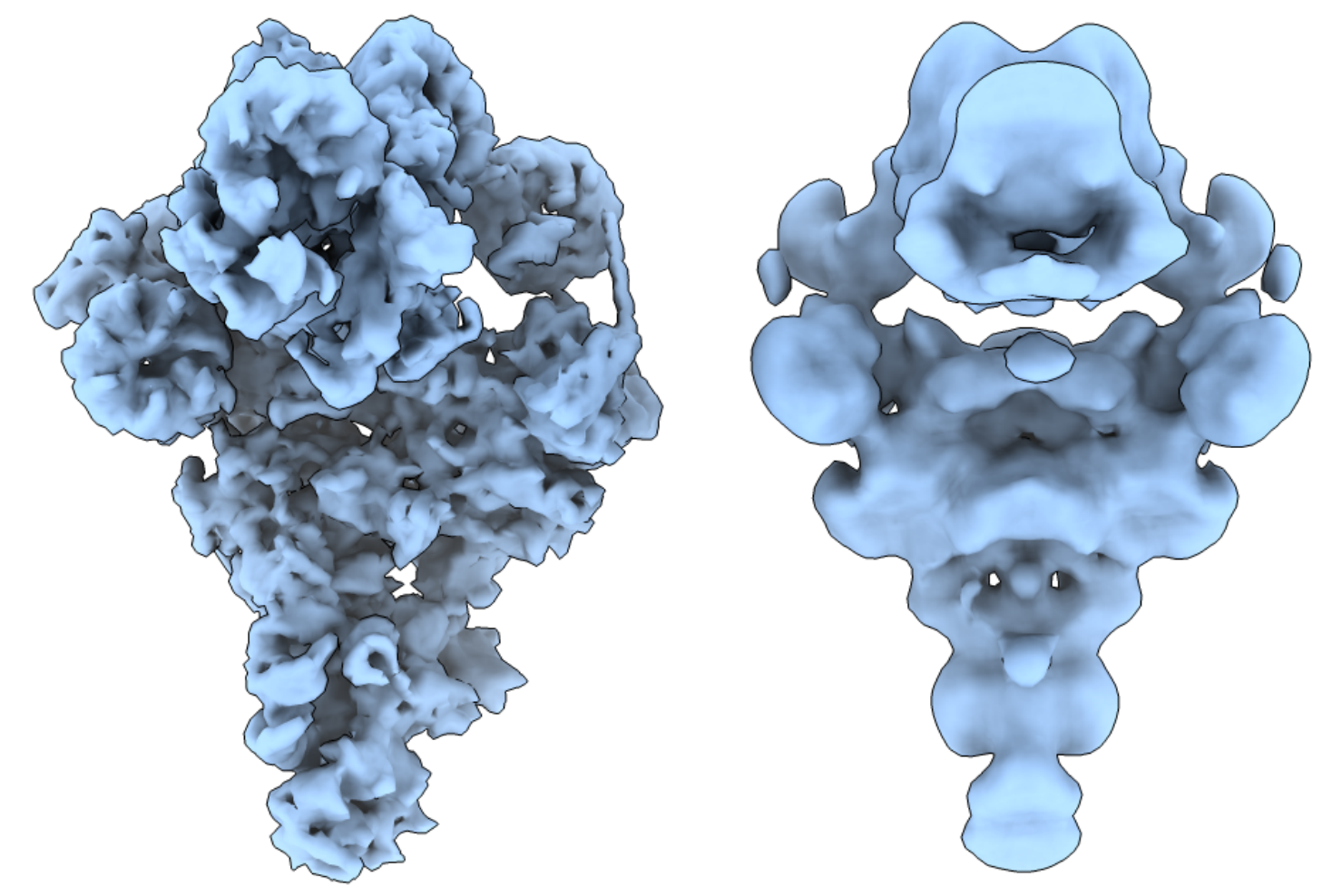}}
  \subcaptionbox*{80S ribosome structure}[.47\linewidth]{%
    \centering 
    \includegraphics[width=0.8\linewidth]{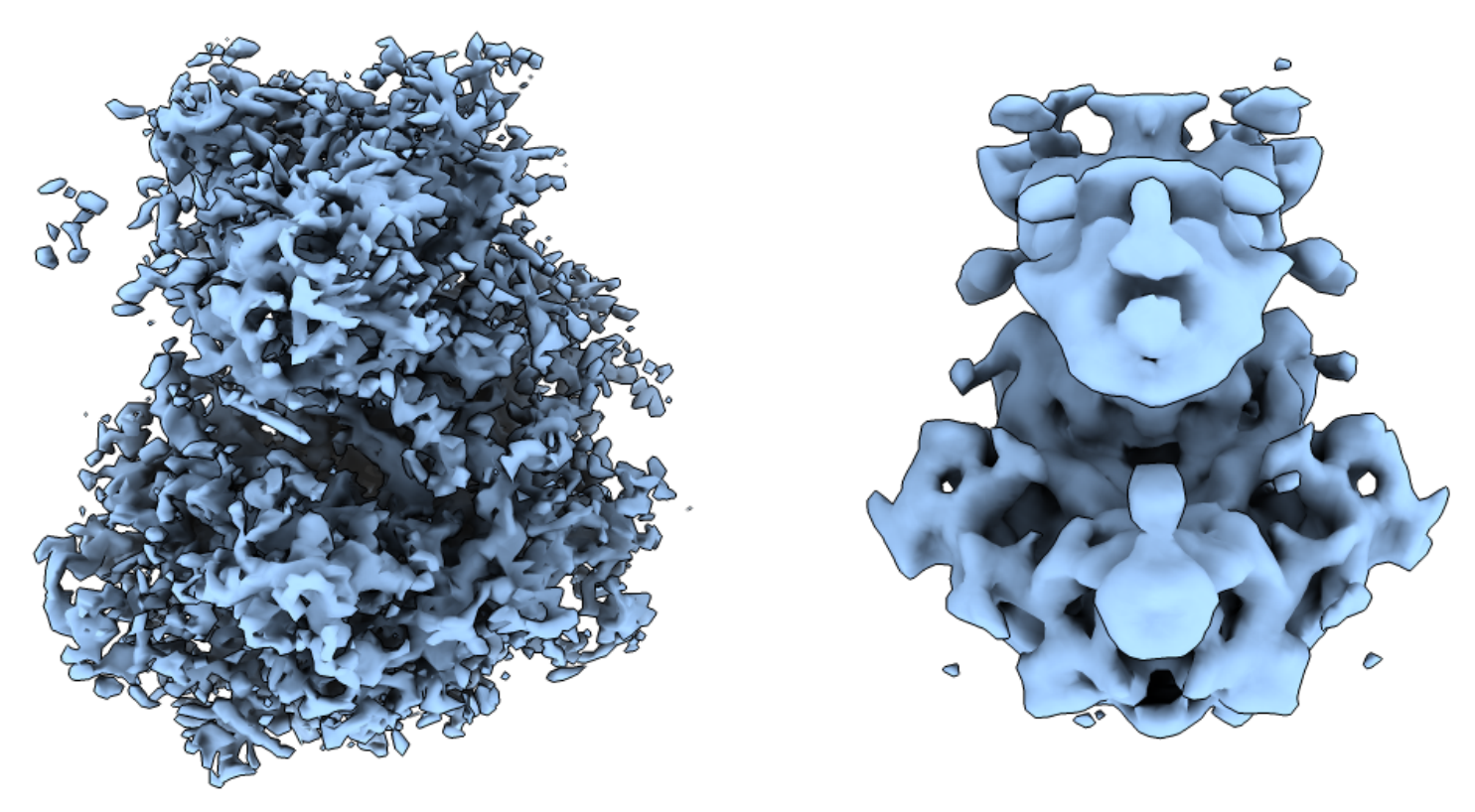}}
  \caption{Visualization of ground truth volume and reconstructed volume showing spurious planar symmetry for the spliceosome structure (left) and the 80S ribosome structure (right). For both structures, the ground truth volume is shown on the left, and the reconstructed volume on the right. Visualizations were made using ChimeraX \cite{meng2023ucsf}}
  \label{spsymm}
\end{figure}

\paragraph{Spurious planar symmetries} In addition, the discussed states corresponded to the decoder reconstructing volumes that have a spurious planar symmetry. In Figure \ref{spsymm} we show reconstructions with such symmetries together with to their corresponding ground truth volumes. This phenomenon that is also discussed in \cite{levy2022cryoai}, in which it was shown that a symmetric loss function prevents the model from getting stuck in states showing these symmetries. Our exploration shows that even with a symmetric loss function, these states occur.

Upon comparing the reconstructions showing planar symmetry with the projective plane visualizations as shown in Figure \ref{pp80s} and Figure \ref{ppspl}, we find that for the patterns in which a subset of points is mirrored with respect to a line in the xy-plane, this line corresponds to the plane of symmetry of the reconstructed volume.

\begin{figure}[h!]
  \centering
      \captionsetup[subfigure]{justification=centering}

    \subcaptionbox*{}[0.49\linewidth]{
  \centering
  \subcaptionbox*{Reconstructed volume}[\linewidth]{%
    \centering 
    \includegraphics[width=\linewidth]{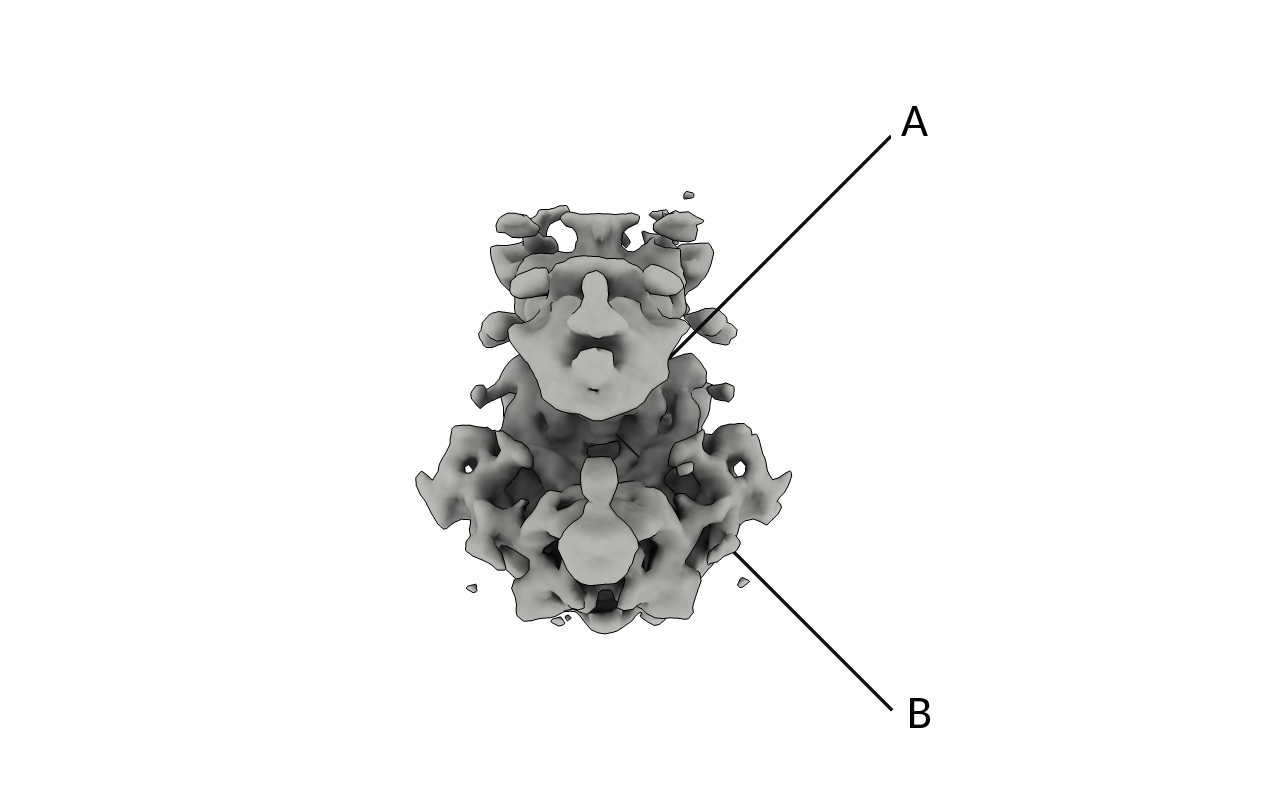}%
  }
  \\
  \subcaptionbox*{Projection with viewing direction (A).}[0.45\linewidth]{%
    \centering 
    \includegraphics[width=\linewidth]{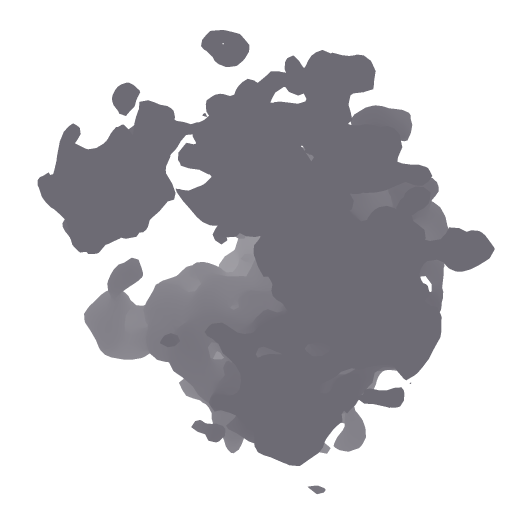}%
  }
    \subcaptionbox*{Projection with viewing direction (B).}[0.45\linewidth]{%
    \centering 
    \includegraphics[width=\linewidth]{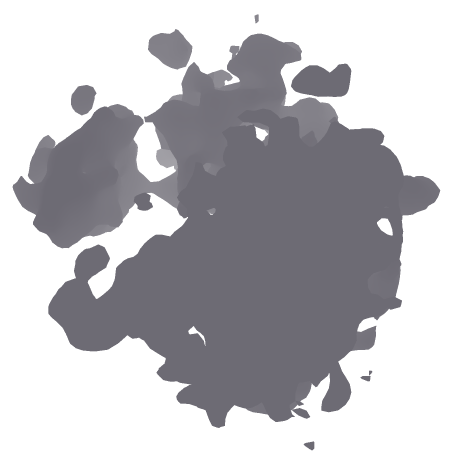}%
  }}
    \subcaptionbox*{}[0.49\linewidth]{
      \centering
  \subcaptionbox*{Ground truth volume}[\linewidth]{%
    \centering 
    \includegraphics[width=0.92\linewidth]{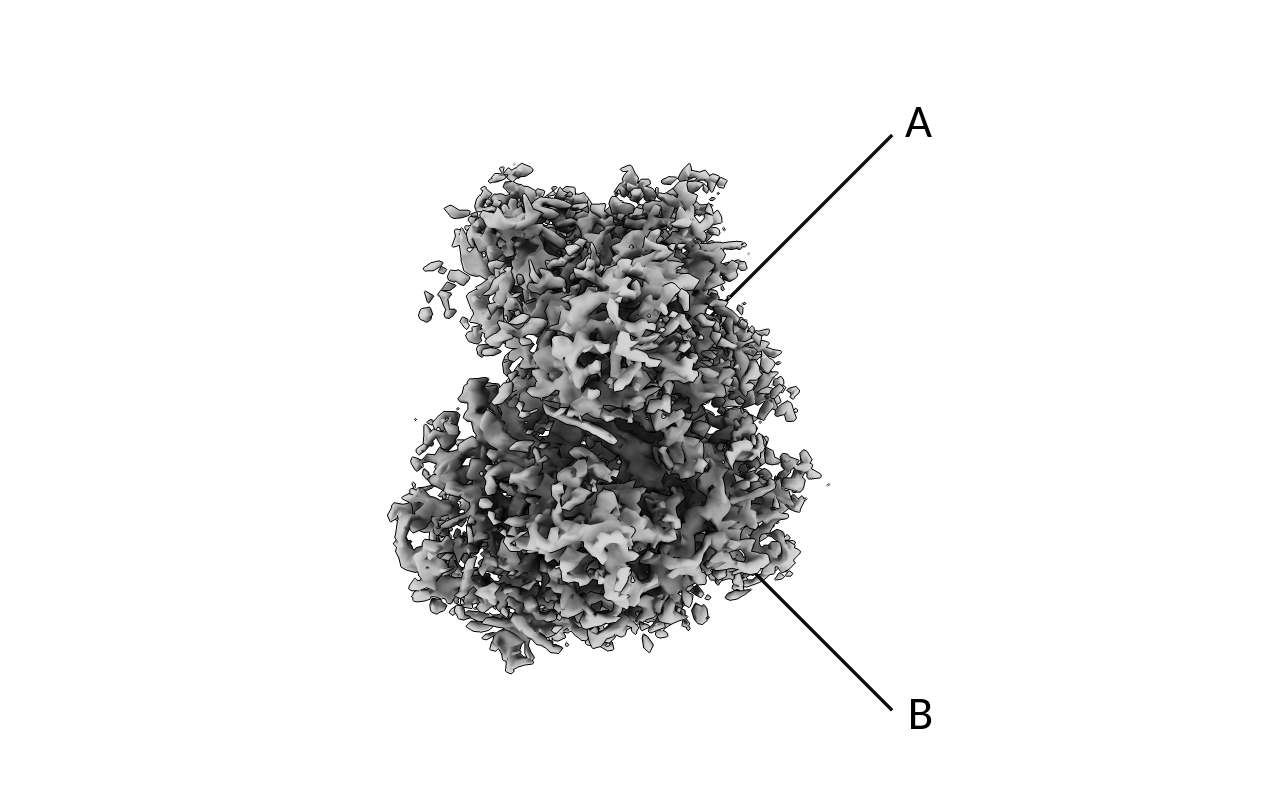}%
  }
  \\
  \subcaptionbox*{Projection with viewing direction (A).}[0.45\linewidth]{%
    \centering 
    \includegraphics[width=\linewidth]{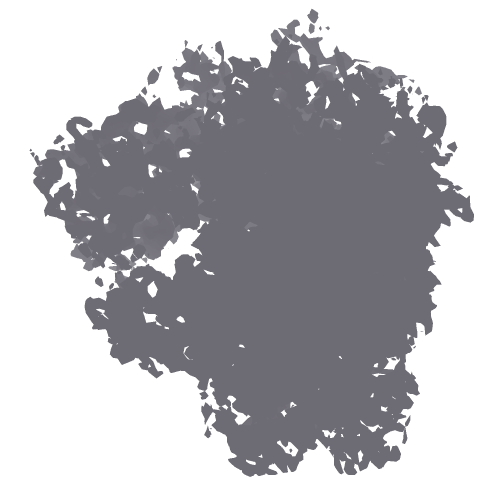}%
  }
    \subcaptionbox*{Projection with viewing direction (B).}[0.45\linewidth]{%
    \centering 
    \includegraphics[width=\linewidth]{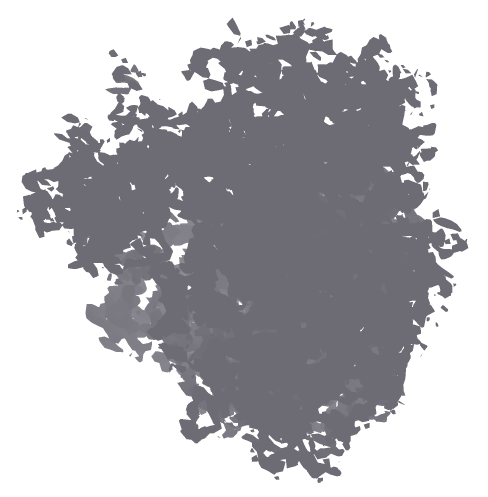}%
  }}
  \caption{A reconstructed volume showing planar symmetry with two viewing directions (A) and (B) is shown on the top left. Viewing direction (B) is the viewing direction that is obtained if (A) is mirrored along the plane of symmetry and multiplied by $-1$. On the top right, the ground truth volume is shown with the same viewing directions. For the volume showing planar symmetry, viewing directions (A) and (B) generate projections that look almost identical due to planar symmetry. On the other hand, the ground truth volume does not show planar symmetry and the projections generated by both viewing directions look different. Visualizations were made with ChimeraX \cite{meng2023ucsf}.}
  \label{concept}
\end{figure}

We hypothesize that the patterns in the projective plane and the planar symmetry in the reconstructed volume can be explained by the fact that for volumes that have a planar symmetry, some viewing directions yield very similar projections, even though they would yield different projections in a non-symmetric volume. We show an example of this in Figure \ref{concept}. This enables the pipeline to reconstruct images in the dataset with a low loss in two disctinct ways. The pipeline in stuck in a state reconstructing a volume with planar symmetry when the encoder is equally likely to predict each of these two. This explains the two equally sized internally consistent subsets of poses.

\paragraph{Interpretation of results} The cryoAI authors (\cite{levy2022cryoai}) have noted that without a symmetric loss function, the pipeline could remain in local minima featuring reconstruction showing spurious planar symmetries for extended periods. We find that even with a symmetric loss function, the standard model can temporarily get stuck in similar local minima. In these states, due to planar symmetry, two distinct poses can generate the same projection, unlike in non-symmetric volumes. The exploratory study also reveals that when the model is in a state reconstructing a symmetric volume, the poses predicted by the encoder form two equally sized groups, each internally consistent but inconsistent with each other. We hypothesize that these groups represent the two distinct ways of generating the same projection in a  volume showing spurious planar symmetry.

We hypothesize that the symmetric loss helps the model escape states in which it is reconstructing volumes with spurious planar symmetries as follows. Utilizing a symmetric loss function involves augmenting each batch with a rotated (or in the case of a mirror loss, mirrored) duplicate of each image in the batch. Both versions of the image pass through the model, but the model is only supervised on the best reconstruction. Recall that we observed that, when reconstructing a volume showing a spurious planar symmetry, two distinct poses can generate the same projection. We expect that at least sometimes, rotating (or mirroring) an image causes the encoder to switch from predicting one pose to the other that can generate the same projection.  If, by chance, one aspect of the symmetry in the reconstructed volume provides marginally better reconstructions for a specific image, the model may begin to prefer this aspect, and as such `escape' the symmetric state.

\end{document}